\documentclass[12pt,a4paper,american,aps,prl,superscriptaddress,reprint]{revtex4-1}
\usepackage{mathptmx}
\usepackage{helvet}
\usepackage{courier}

\usepackage[T1]{fontenc}
\usepackage[utf8]{inputenc}
\usepackage{color}
\usepackage{babel}
\usepackage{amsmath}
\usepackage{amssymb}
\usepackage{graphicx}
\usepackage[unicode=true,pdfusetitle,
 bookmarks=true,bookmarksnumbered=true,bookmarksopen=true,bookmarksopenlevel=2,
 breaklinks=true,pdfborder={0 0 1},backref=false,colorlinks=true]
 {hyperref}
\hypersetup{
 allcolors=blue}

\makeatletter

\pdfpageheight\paperheight
\pdfpagewidth\paperwidth

\providecommand{\tabularnewline}{\\}

\makeatother

\begin{document}
\title{Dynamics of hydrogen in silicon at finite temperatures from first-principles}
\author{Diana Gomes}
\address{I3N, Department of Physics, University of Aveiro, Campus Santiago,
3810-193 Aveiro, Portugal}
\author{Vladimir P. Markevich}
\address{Photon Science Institute and Department of Electrical and Electronic
Engineering, The University of Manchester, Manchester M13 9PL, United
Kingdom}
\author{Anthony R. Peaker}
\address{Photon Science Institute and Department of Electrical and Electronic
Engineering, The University of Manchester, Manchester M13 9PL, United
Kingdom}
\author{Jos\'{e} Coutinho}
\email{jose.coutinho@ua.pt}

\address{I3N, Department of Physics, University of Aveiro, Campus Santiago,
3810-193 Aveiro, Portugal}
\begin{abstract}
Hydrogen point defects in silicon still hold unsolved problems, whose
disclosure is fundamental for future advances in Si technologies.
Among the open issues is the mechanism for the condensation of atomic
hydrogen into molecules in Si quenched from above $T\sim700$~$^{\circ}$C
to room temperature. Based on first-principles calculations, we investigated
the thermodynamics of hydrogen monomers and dimers at finite temperatures
within the harmonic approximation. The free energies of formation
indicate that the population of {\normalsize{}H$^{-}$} cannot be
neglected when compared that of {\normalsize{}H$^{+}$} at high temperatures.
The results allow us to propose that the formation of molecules occurs
during cooling processes, in the temperature window $T\sim700\textrm{-}500$~K,
above which the molecules collide with Si-Si bonds and dissociate,
and below which the fraction of {\normalsize{}H$^{-}$} becomes negligible.
The formation of {\normalsize{}H$^{-}$} and most notably of a fast-diffusing
neutral species could also provide an explanation for the apparent
\emph{accelerated} diffusivity of atomic hydrogen at elevated temperatures
in comparison to the figures extrapolated from measurements carried
out at cryogenic temperatures. We finally show that the observed diffusivity
of the molecules is better described upon the assumption that they
are nearly free rotors, all along the minimum energy path, including
at the transition state.\emph{ {[}Pre-print published in Physica Status Solidi B {\bfseries 2022}, 2100670. Invited by the occasion of the 60th anniversary of
the journal{]}}

\noindent \href{https://doi.org/10.1002/pssb.202100670}{DOI:10.1002/pssb.202100670}
\end{abstract}

\maketitle
\noindent 

\section{Introduction}

The presence of hydrogen in crystalline silicon is virtually unavoidable.
H species are highly mobile and reactive, they can change several
properties of the host upon interactions with point and extended defects,
as well as with contaminants and dopants. Hydrogen is employed on
wafer processing and surface treatments, it is determinant in the
performance of oxide-semiconductor interfaces, and for these and many
other reasons, few are the elements that deserved so much attention
in the past 60 years or so of research of defects in silicon \citep{Stutzmann1991,Pankove1991,Pearton1992,Nickel1999}.
The current understanding of hydrogen point defects in Si has been
recently reviewed in Refs.~\citep{Peaker2008,Estreicher2014}, both
from the experimental and theoretical perspectives. We only summarize
the essential features for the sake of the present work.

From electron paramagnetic resonance \citep{Nielsen1994,Gorelkinskii1996,Ammerlaan2002},
deep level transient spectroscopy (DLTS) \citep{Johnson1994,Nielsen1999},
Fourier transform infra-red spectroscopy \citep{Budde2000}, muon
spin rotation spectroscopy \citep{Hitti1999}, and theoretical modeling
\citep{Sasaki1989,Walle1989,Jones1991,Herring2001,Hourahine2003,Estreicher2012},
there is ample evidence and reasoning for the presence of isolated
hydrogen atoms and ions in Si. Atomic H can trap a hole to become
a proton located at the \emph{bond-center} site (H$_{BC}^{+}$). It
can also trap one electron to become a hydride ion which, according
to theorists, finds its most stable state in the open regions of the
lattice, either at the \emph{tetrahedral} interstitial site (H$_{T}^{-}$)
\citep{Walle1989,Herring2001,Estreicher2012}, or close to it, at
the \emph{anti-bonding} site (H$_{AB}^{-}$) \citep{Sasaki1989,Jones1991,Karazhanov2014}.
This ambiguity reflects the shallowness of the potential energy for
the roaming of H$^{-}$ within the volume enclosed by the tetrahedral
cage of the Si lattice \citep{Estreicher2012}. The above crystalline
sites are represented in Figure~\ref{fig1}.

Single hydrogen atom in Si shows negative-$U$ properties and bistability
in the neutral charge state \citep{Johnson1994}. Essentially, $\textrm{H}^{0}$
is metastable against disproportionation, $2\textrm{H}^{0}\rightarrow\textrm{H}^{+}+\textrm{H}^{-}$
\citep{Watkins1984,Coutinho2020}. This is clearly understood if we
break the reaction into two steps, and realize that the small ionization
energy for $\textrm{H}_{BC}^{0}\rightarrow\textrm{H}_{BC}^{+}+\textrm{e}^{-}$
($0.175$~eV), is overcompensated by the affinity for $\textrm{H}_{BC}^{0}+\textrm{e}^{-}\rightarrow\textrm{H}_{T}^{-}$
(estimated between $-0.4$~eV and $-0.6$~eV) \citep{Nielsen1999,Nielsen2002}.
Importantly, because only the metastable $\textrm{H}_{T}^{0}$ state
can become negatively charged, the second step includes a reconfiguration
barrier of $\sim0.3$~eV involving $\textrm{H}_{BC}^{0}\rightarrow\textrm{H}_{T}^{0}$
that must be surmounted \citep{Nielsen1999}. This barrier explains
why H at the $BC$ site is the dominant species found in low-temperature
proton-implanted n-type Si \citep{Nielsen1994,Gorelkinskii1996},
despite the fact that under these conditions the ground state is $\textrm{H}_{T}^{-}$
(or $\textrm{H}_{AB}^{-}$). The above picture implies that atomic
H has a $(-/+)$ occupancy level in the range 0.3-0.4~eV below the
conduction band bottom ($E_{\textrm{c}}$), with the uncertainty arising
from difficulties in measuring the energy of $\textrm{H}_{T}^{0}$
with respect to $\textrm{H}_{BC}^{0}$ \citep{Nielsen2002}.

It is also very hard to measure the diffusivity of hydrogen without
the result being affected by impurity trapping (dopants, oxygen and
carbon, to name some of the most important). Two experiments that
avoided this effect are highlighted. In a 1996 paper, the decay rate
of a stress-aligned population of $\textrm{H}_{BC}^{+}$ defects,
was measured during annealing in the temperature range $T=126\textrm{-}142$~K
by Gorelkinskii and Nevinnyi (GN) \citep{Gorelkinskii1996}. The result,
$\nu=(2.3\,\textrm{THz})\exp(-0.43\pm0.02\,\textrm{eV}/k_{\textrm{B}}T)$,
was interpreted as reflecting the jump of individual protons between
neighboring bond-center sites during the return to natural randomness.
The pre-exponential factor is consistent with a phonon activated single
jump. Following Herring et~al. \citep{Herring2001}, we can convert
the measured dichroism decay rate, $\nu$, into diffusivity as $D=l^{2}\nu/8$,
which gives $D=(1.06\times10^{-4}\,\textrm{cm}^{2}/\textrm{s})\exp(-0.43\pm0.02\,\textrm{eV}/k_{\textrm{B}}T)$,
considering the distance between neighboring bond-centered sites $l=1.92$~Å
as the elementary jump length. In the other experiment, published
in 1956, the escape of pressurized molecular hydrogen through thin-walled
Si canisters was monitored by Van Wieringen and Warmoltz (VWW) in
the temperature range $T=1090\textrm{-}1200$~$^{\circ}$C \citep{vanWieringen1956}.
At such high temperatures, trapping processes become insignificant,
and a diffusivity $D=(9.4\times10^{-3}\,\textrm{cm}^{2}/\textrm{s})\exp(-0.48\pm0.05\,\textrm{eV}/k_{\textrm{B}}T)$
was found. From the pressure-dependence of the hydrogen permeation,
it was clear that hydrogen migrated in the atomic form. Considering
the close activation energies extracted from the low and high temperature
data, it is commonly assumed that transport of atomic hydrogen in
Si occurs via consecutive jumps of H$^{+}$ between neighboring bond-center
sites.

Among the many measurements of the hydrogen diffusivity close to room
temperature (see Ref.~\citep{Peaker2008} for a collection of data),
the one of Kamiura, Yoneta and Hashimoto (KYH) \citep{Kamiura1991}
falls distinctively close to the line that connects the VWW and GN
data points. In this work the diffusion-related kinetics of H released
from photo-induced dissociation of carbon-hydrogen complexes near
the surface of the silicon (formed via wet-etching) was measured in
the range $T=220\textrm{-}270$~K, leading to a diffusivity $D=(7\times10^{-2}\,\textrm{cm}^{2}/\textrm{s})\exp(-0.54\,\textrm{eV}/k_{\textrm{B}}T)$.
This result was claimed to reflect a rate-limiting process involving
the diffusion of hydrogen to phosphorous atoms \citep{Kamiura1991}.

Hydrogen in Si is also found in the form of pairs. At least two H
dimers were found experimentally in the Si lattice, namely the hydrogen
molecule located at the tetrahedral interstitial site (H$_{2T}$)
and a close pair of Si-H$_{BC}\cdots$Si-H$_{AB}$ units, referred
to as H$_{2}^{*}$. While in very pure Si the molecule can represent
the main stock of free hydrogen available at room temperature (e.g.
Ref.~\citep{Binns1993}), H$_{2}^{*}$ has been mostly detected in
irradiated material \citep{Holbech1993}, despite having a formation
energy of only few tenths of eV higher than the molecule \citep{Estreicher2004}.
Formation of H$_{2}^{*}$ in non-irradiated Si has also been reported
by Suesawa though \citep{Suezawa1999}. A possible route for its formation
has been proposed by Estreicher \emph{et~al.}\citep{Estreicher1999},
and involves the interaction of H$_{2}$ molecules with radiation
induced defects. From a re-interpretation of the out-diffusion kinetics
of tritium quenched samples \citep{Ichimiya1968}, the existence of
a third and non-detectable hydrogen dimer, referred to as $H_{2\textrm{B}}$,
has been postulated by Voronkov and Falster \citep{Voronkov2017a}.
This species would account for the difference in the concentration
of hydrogen-boron complexes (HB) formed after complete conversion
of H$_{2T}$ molecules into HB in boron doped material annealed at
160~$^{\circ}$C and 175~$^{\circ}$C. Assuming that $H_{2\textrm{B}}$
is not stable at the higher temperature, its dissociation would explain
the observation of a 45\% increase in the formation of HB pairs.

Although being electrically inert, H$_{2T}$ molecules can be detected
by local vibrational mode spectroscopy, either via Raman scattering
or absorption in the infra-red region \citep{Holbech1993,Leitch1998,Pritchard1998}.
They can be introduced in Si upon exposing the Si to a low temperature
($\sim$150~$^{\circ}$C) hydrogen plasma \citep{Leitch1998} or
to a high temperature ($T\gtrsim700$~$^{\circ}$C) gas phase followed
by quenching. In the latter case, concentrations of interstitial molecules
of a few times $10^{15}$~cm$^{-3}$ could be reached \citep{Pritchard1998}.
The molecules become mobile above 30~$^{\circ}$C, and can be trapped
by/near impurities such as interstitial oxygen (O$_{\textrm{i}}$)
\citep{Markevich1998}, substitutional carbon \citep{Peng2011}, or
substitutional boron \citep{Pritchard1999}. From the dissociation/recovery
kinetics of O$_{\textrm{i}}$-H$_{2}$, a diffusivity $D=(2.6\pm1.5\,\textrm{cm}^{2}/\textrm{s})\times10^{-4}\exp(-0.78\mp0.05~\textrm{eV}/k_{\textrm{B}}T)$
was attributed to the migration of H$_{2}$ across the Si lattice
\citep{Markevich1998}.

Despite the success in predicting and describing the properties of
H-related point defects by first-principles methods, including their
vibrational and electronic properties (see for instance Refs.~\citep{Andersen2002,Pruneda2002,Hourahine1998,Peng2011}),
several issues were left unsolved. Among the fundamental problems,
the most intriguing is perhaps the mechanism for the formation of
H$_{2}$ molecules from atomic hydrogen upon cooling the crystals
from high temperatures. The main difficulty is that the Fermi level
is close to mid gap and with the $(-/+)$ level of hydrogen at 0.3-0.4~eV
below $E_{\textrm{c}}$, the population of H monomers is thought to
consist essentially of mutually repelling H$^{+}$ ions.

The migration of hydrogen also has open issues. As pointed out in
Ref.~\citep{Voronkov2017b}, if we extrapolate the diffusivity from
the low temperature data to the range of the high temperature experiment,
the diffusivity obtained for H$^{+}$ is several times lower than
that recorded by VWW. It was then suggested that neutral hydrogen,
although present in small quantities, could dominate the diffusivity
at high temperatures due to its low migration barrier \citep{Voronkov2017a,Voronkov2017b}.
Despite its significance, this observation is based on a rather extreme
extrapolation spanning 17 orders of magnitude of diffusivity, between
the high temperature data of VWW \citep{vanWieringen1956}, which
covered a narrow range of only a factor of two, and the low temperature
data of GN \citep{Gorelkinskii1996} obtained over two decades.

Another issue is the migration barrier of H$^{-}$ which has been
calculated as 0.39~eV \citep{Estreicher2012}, nearly half of the
value (0.7~eV) that was obtained from the kinetics of hydrogen passivation/reactivation
of phosphorous donors \citep{Johnson1992}. A possible solution for
this discrepancy involves an alternative interpretation of the measured
0.7~eV barrier, which would correspond to a reorientation of the
P-H complex accompanied by a charge state change \citep{Estreicher1991}.
The neutral state is even more puzzling — the barrier for migration
between $\textrm{H}_{BC}^{0}$ ground states has a calculated value
of 0.38~eV \citep{Estreicher2012}. No measurements are available
for this figure, essentially because of the short lifetime of neutral
hydrogen. A striking observation was that upon forward-bias injection
of p$^{+}$-n diodes at 65~K \citep{Nielsen1999}, when most hydrogen
atoms were assumed to be in the metastable neutral state (referred
to as H$_{T}^{0}$), the estimated diffusivity was 27 orders of magnitude
higher compared to that of H$^{+}$ at the same temperature (as extrapolated
from the VWW data \citep{vanWieringen1956}). If we accept such an
extrapolation as meaningful (based on data that spans less than a
decade of diffusivity), it suggests that there is a barrier preventing
the relaxation $\textrm{H}{}_{T}^{0}\rightarrow\textrm{H}{}_{BC}^{0}$,
which must be higher than the barrier for migration between $\textrm{H}{}_{T}^{0}$
states, and the latter should be lower than 0.1~eV \citep{Nielsen2002}.

A rather impactful problem involving hydrogen in silicon is light-
and elevated temperature-induced degradation (LeTID) of silicon solar
cells \citep{Ramspeck2012}. This effect is responsible for a decrease
of the power conversion efficiency of modules by up to 16\% relative,
and it was found to be particularly detrimental in new-generation
multicrystalline passivated emitter and rear cells (see Ref.~\citep{Chen2020}
for a recent review). Based on junction spectroscopy and first-principles
calculations, the relocation of hydrogen and the formation of a boron-dihydride
complex (BH$_{2}$) in p-type Si has been proposed as a possible culprit
for LeTID \citep{Guzman2021}.

The understanding of LeTID is strongly tied with our knowledge regarding
the issues discussed above. They involve processes driven by kinetics
and excitations (quenching, illumination, annealing), all of which
cannot be understood if we leave vibrational and electronic excitations
out of the physical picture. The recent work by Sun and co-workers
\citep{Sun2015,Sun2021} revised the second (electronic) type of excitations,
with an evaluation the relative concentrations of different charge
states of H under non-equilibrium steady-state carrier injection.
In this paper, we consider the effect of vibrational degrees of freedom
on several hydrogen related properties and processes.

The next sections are organized in the following manner: Section~\ref{sec:method}
describes the methodology, including the calculation of electronic
plus clamped ion energies, free energies of formation, activation
energy barriers, reaction rates and diffusion coefficients. The bulk
of the results are described in Section~\ref{sec:results}. Along
the way, we will (1) present a fresh look into the configuration coordinate
diagram of atomic H, describing carrier trapping/emission, as well
as transformation and migration processes; (2) discuss a mechanism
for the formation of molecules in quenched samples; (3) reconcile
the low temperature measurements of the jump rate of $\textrm{H}{}_{\textrm{BC}}^{+}$
with the high-temperature diffusivity. The paper ends with a table
summarizing the main results and assignments along with several concluding
remarks.

\section{Methodology\label{sec:method}}

\subsection{All-electron energies\label{subsec:ei-energies}}

All-electron energies were calculated within density functional theory
employing the plane-wave pseudopotential formalism \citep{Kresse1993,Kresse1994,Kresse1996a,Kresse1996b}.
Either a semi-local density functional (generalized gradient approximated,
GGA \citep{Perdew1996}) or a range-separated hybrid functional (as
proposed by Heyd, Scuseria, and Ernzerhof, HSE \citep{Heyd2003,Heyd2006})
described the electronic exchange-correlation interactions. The projector-augmented
wave method was used to treat the core electrons \citep{Bloechl1994},
whereas valence states were described by plane-waves. The self-consistent
electron density and potential were converged until the total energy
between two consecutive steps differed by less than $10^{-8}$~eV.

\begin{figure}
\includegraphics{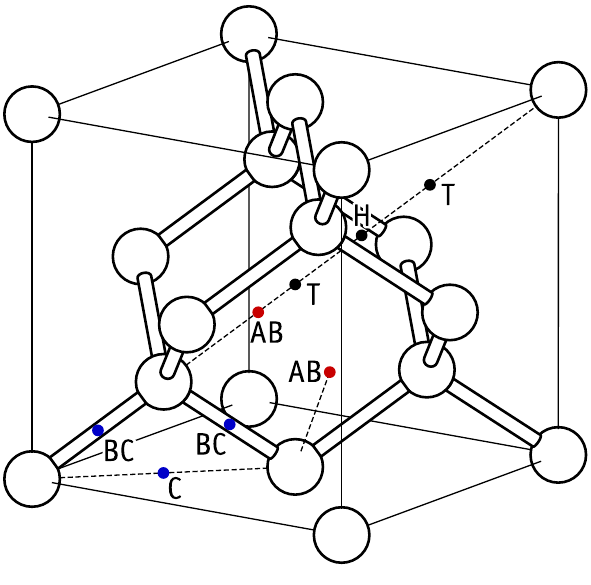}

\caption{\label{fig1}Conventional cell of silicon. High symmetry crystalline
sites are indicated with dots along with (commonly used) labels. The
coloring of dots is helpful for the discussion of migration and reorientation
mechanisms of hydrogen defects in different charge states.}
\end{figure}

Several hydrogen defects were investigated. As for hydrogen monomers,
the sites investigated are identified in the conventional unit cell
of Si depicted in Figure~\ref{fig1}. The usual labels are used for
the bond-centered, anti-bonding, tetrahedral and hexagonal ($H$)
interstitial sites. Site $C$ is also included for the sake of discussing
migration and reorientation mechanisms. The different coloring of
the dots also serves that purpose.

Two kinds of hydrogen dimers were investigated with particular detail,
namely the molecule at the $T$ site, H$_{2T}$, and the H$_{2}^{*}$
complex. The latter is made of neighboring H$_{\textrm{BC}}$ and
H$_{\textrm{AB}}$ along the principal diagonal of the cube represented
as a dashed line in Figure~\ref{fig1}.

\noindent 

Three batches of first-principles calculations were carried out, namely
(i) total energy of stable structures; (ii) minimum energy paths and
respective energy barriers between stable structures; (iii) vibrational
mode frequencies.

In type (i) calculations the stable geometries of defects and respective
energies were found in cubic supercells of 512 Si atoms with the plane-wave
cut-off energy set to $E_{\textrm{cut}}=400$~eV. The Kohn-Sham problem
was solved at $\mathbf{k}=\Gamma$ only. Structural optimization (within
GGA-level) ensured that the residual forces acting on the nuclei were
lower than 0.01~eV/Å. A self-consistent calculation within HSE-level
was performed to evaluate the electronic energy of the periodic supercell,
$\tilde{E}_{\textrm{elec}}$. Finally, a periodic charge correction
was added to $\tilde{E}_{\textrm{elec}}$ in order to find the (clamped-ion)
electronic energy of the defective cell, $E_{\textrm{elec}}=\tilde{E}_{\textrm{elec}}+E_{\textrm{corr}}$
\citep{Freysoldt2009}.

In type (ii) calculations, defect migration, reorientation and transformation
mechanisms were investigated by finding minimum energy paths using
the climbing image nudged elastic band (NEB) method \citep{Henkelman2000}.
Accordingly, 11 intermediate structures between the initial and final
states were relaxed while subject to the \emph{elastic band} constraint.
For that, we employed 64-atom cubic cells, $E_{\textrm{cut}}=400$~eV,
a grid of 2x2x2 $\mathbf{k}$-points for sampling the Brillouin zone,
and the GGA exchange-correlation functional. Hybrid density functional
single-point calculations were finally carried out in order to find
the electronic component for the activation energy barrier $\Delta E_{\textrm{a,elec}}=E_{\textrm{elec}}^{\ddagger}-E_{\textrm{elec}}$,
where $E_{\textrm{elec}}^{\ddagger}$ and $E_{\textrm{elec}}$ are
HSE-level energies of the transition state and initial state, respectively.

In type (iii) calculations we constructed the dynamical matrix from
the force constants $\Phi_{ij}=-\partial F_{i}/\partial x_{j}$ using
numerical differentiation with atomic displacements $\Delta x=0.01$~Å.
Here $\{i,j\}=1,\ldots,3N$, $F_{i}$ is a force component along $i$,
$x_{j}$ an atomic displacement along $j$ and $N$ the total number
of atoms in the supercell. The dynamical matrix elements were found
at $\mathbf{q}=\Gamma$. Harmonic vibrational mode frequencies were
found via matrix diagonalization. These calculations were performed
at the GGA-level with $E_{\textrm{cut}}=500$~eV, the respective
equilibrium structures were found in $N=64$ atom supercells, residual
forces of the relaxed structures were lower than 0.005\,eV/Å, and
a grid of $2\times2\times2$ $\mathbf{k}$-points was used for BZ
integration \citep{Monkhorst1976}. Convergence of the vibrational
properties due to finite size effects were verified using larger 216
atom cells. Further testing has been reported elsewhere \citep{Estreicher2004,Murali2015}.

\subsection{Free energies of formation}

The vast majority of experiments related to defects in silicon is
performed at constant-pressure. Therefore, quantities that are probed
under thermodynamic equilibrium conditions, relate to the (change
in the) Gibbs free energy, $\Delta G$. On the other hand, from the
perspective of modeling, it is convenient to work under constant-volume
and calculate changes in the Helmholtz free energy, $\Delta F$ (see
for instance \citep{AlMushadani2003,Zhang2018}). While for sufficiently
large supercells, the potential contributions to the free energy (stationary
ionic and electronic solutions) converge to the same value under constant-pressure
and constant-volume regimes \citep{Mishin2001}, the same is not true
for the vibrational free energy, which being affected by anharmonicity,
induces a ($T$-dependent) macroscopic volume change to the crystal.
However, as pointed out in Ref.~\citep{Estreicher2004}, the minute
thermal expansion of crystalline Si indicates that phonon frequencies
change weakly with $T$, and that constant-volume and constant-pressure
calculations are comparable up to a few hundred degrees Celsius. This
is an important observation which is also assumed by us, although
it should be taken with due care by the reader.

Regarding the calculation of the free energy of formation of defects
in crystals, the formalism can be found elsewhere \citep{Estreicher2004,Huang2010,Murali2015}.
Here we leave a summary of the main ingredients and approximations
considered. Our starting point is a sample made of $N_{\textrm{L}}$
lattice sites spanning, a constant volume $V$, and containing $n_{\textrm{d}}$
defects. We assume a dilute limit, that is, the concentration of defects
under scrutiny is low enough so that we can (1) ignore defect-defect
interactions, including in the calculation of configurational entropy,
and (2) neglect their impact on the location of the Fermi level. Accordingly,
we can define the Helmholtz free energy of formation of a defective
sample as,

\begin{equation}
\Delta F_{\textrm{f}}=F-F^{(0)},\label{eq:free_form}
\end{equation}
where $F$ is the total free energy enclosed in the volume $V$, which
includes potential terms as well as temperature-dependent excitations
and entropy. Here we consider electronic, vibrational, rotational
(for the case of molecular H$_{2}$ in Si) and configurational degrees
of freedom,

\begin{equation}
F=F_{\textrm{elec}}+F_{\textrm{vib}}+F_{\textrm{rot}}-TS_{\textrm{conf}}.\label{eq:free_tot}
\end{equation}

The quantity $F^{(0)}$ is a reference free energy which is further
detailed below. The first term in Eq.~\ref{eq:free_tot} accounts
for the electronic (clamped-ion) internal energy and entropy, $F_{\textrm{elec}}=U_{\textrm{elec}}-TS_{\textrm{elec}}$.
However, for defects with deep gap states, $F_{\textrm{elec}}$ is
usually replaced by the zero-temperature internal energy, $U_{\textrm{elec}}(T=0\,\textrm{K})=E_{\textrm{elec}}$.
The reason is that (1) the fraction of defects that can be promoted
to electronic excited states drops exponentially with the excitation
energy, and (2) the electronic entropy is proportional to the electronic
density of states within $E_{\textrm{F}}\pm k_{\textrm{B}}T$, where
$E_{\textrm{F}}$ and $k_{\textrm{B}}$ are the Fermi level and the
Boltzmann constant, respectively \citep{Ashcroft1976}.

Estreicher \emph{et~al.} \citep{Estreicher2004} showed that the
electronic entropy of defects with states deeper than $\sim0.1$~eV
from the band edges, contributes no more than few tens of meV to the
free energy, even for defect concentrations as high as $10^{17}\,\textrm{cm}^{-3}$
at $T\sim900$~K. Considering that H monomers are the only electronically
active defects being investigated, and that the shallowest state to
be considered is at $\sim E_{\textrm{c}}-0.2$~eV, we can safely
assume that $F_{\textrm{elec}}=E_{\textrm{elec}}$, which is the ground
state energy of the all-electron (plus nuclear) stationary solution
of the problem, here obtained from a hybrid functional calculation
as described in Sec.~\ref{subsec:ei-energies}.

The vibrational free energy of $3N-3$ independent oscillators with
frequency $\omega_{i}$ involving $N$ atoms is,

\begin{equation}
F_{\textrm{vib}}=k_{\textrm{B}}T\sum_{i=1}^{3N-3}\log\left[2\sinh\left(\frac{\hslash\omega_{i}}{2k_{\textrm{B}}T}\right)\right],\label{eq:free_vib}
\end{equation}
where $\hslash$ is the reduced Plank constant. Eq.~\ref{eq:free_vib}
already includes the zero-point vibrational energy, $E_{\textrm{ZP}}=\sum_{i}^{3N-3}\hslash\omega_{i}/2$,
which along with the electronic component defines the potential energy
of the problem $E=E_{\textrm{elec}}+E_{\textrm{ZP}}$. The vibrational
frequencies, $\omega_{i}$, were obtained as described at the end
of Sec.~\ref{subsec:ei-energies}. The vibrational entropy and specific
heat (at constant volume) are readily found from,

\begin{equation}
S_{\textrm{vib}}=-\frac{\partial F_{\textrm{vib}}}{\partial T};\quad c_{\textrm{v}}=-T\left(\frac{\partial^{2}F_{\textrm{vib}}}{\partial T^{2}}\right).\label{eq:s-cv}
\end{equation}

\begin{figure}
\includegraphics{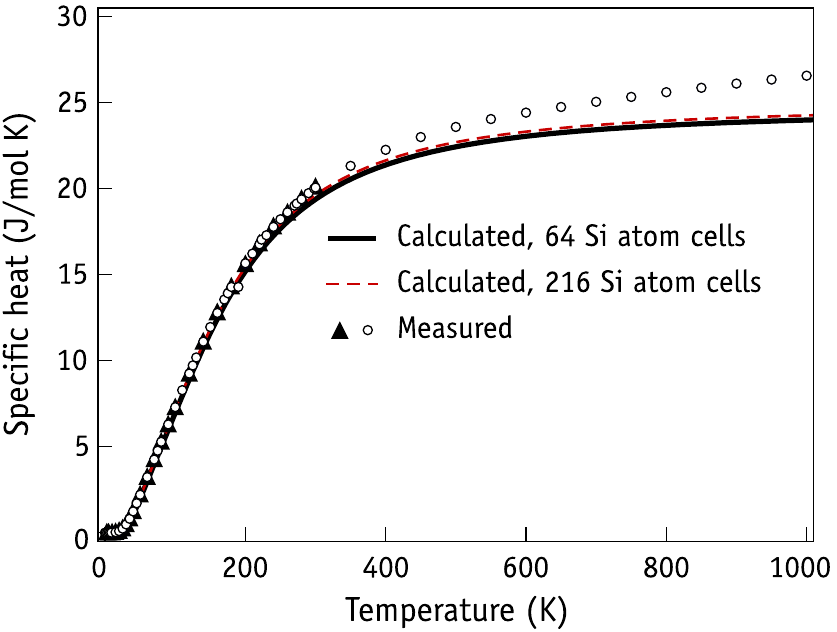}

\caption{\label{fig2}Specific heat of silicon. Circles and triangles represent
data measured at constant pressure conditions, $c_{\textrm{p}}$,
and reported in Refs.~\citep{Flubacher1959,Desai1986}. The two lines
represent the calculated specific heat at constant volume, $c_{\textrm{v}}$,
within the harmonic approximation (Eqs.~\ref{eq:free_vib} and \ref{eq:s-cv})
using 189 vibrational frequencies from a 64-atom supercell (black
thick line), and 645 vibrational frequencies from a 216-atom supercell
(red dashed thin line).}
\end{figure}

Figure~\ref{fig2} depicts the calculated specific heat at constant
volume of bulk silicon, obtained from 189 vibrational frequencies
of a 64-atom supercell (thick black line). In the same graph we reproduce
specific heat data measured at constant pressure \citep{Flubacher1959,Desai1986}.
Clearly the calculation describes the measurements reasonably well
up to a few hundred degrees K. This provides us an indication of the
temperature beyond which the harmonic approximation starts to break.
Above 400-600~K, anharmonic effects become sizable, \emph{i.e.} for
each vibrational mode, the energy separation between consecutive excited
states decreases with increasing the temperature. Hence, beyond these
threshold, finite temperature calculations become more qualitative.

Also shown in Figure~\ref{fig2} is the specific heat calculated
using a cubic supercell made of 216 Si atoms, corresponding to $3N-3=645$
vibrational frequencies (red dashed line). Clearly, the difference
from the smaller 64-atom cell is not substantial, suggesting that
the sampling of the vibrational structure cannot be significantly
improved by increasing the cell size.

\noindent 

In the dilute limit, all contributions to the free energy are extensive
with respect to the number of defects. However, the configurational
entropy, $S_{\textrm{conf}}$, is a non-local quantity that must be
obtained from combinatorial analysis of all defect configurations
across the whole sample volume. Accordingly, the number of distinct
ways a population of $n_{\textrm{d}}$ identical defects can be distributed
over $N_{\textrm{L}}$ lattice sites is,

\begin{equation}
W=\frac{g^{n_{\textrm{d}}}N_{\textrm{L}}!}{(N_{\textrm{L}}-n_{\textrm{d}})!n_{\textrm{d}}!}\approx\frac{(gN_{\textrm{L}})^{n_{\textrm{d}}}}{n_{\textrm{d}}!},\label{eq:W}
\end{equation}
where $g$ is the number of degenerate orientations that each defect
can have per lattice site. From the definition of entropy, $S=k_{\textrm{B}}\log W$,
we can obtain an approximate expression for $S_{\textrm{conf}}$ \emph{per
defect} after using a first order Taylor's series for $\mathrm{e}^{x}$
and Stirling's formula,

\begin{equation}
S_{\textrm{conf}}=k_{\textrm{B}}\left(1-\log\frac{c}{g}\right),
\end{equation}
where $c=n_{\textrm{d}}/N_{\textrm{L}}$ is the defect concentration.

Additional degrees of freedom can be added to Eq.~\ref{eq:free_tot},
depending on the specificity of the problem at hand. For instance,
molecular hydrogen in Si essentially behaves as a free rigid rotor
in the tetrahedral interstitial site \citep{Estreicher2001}. To account
for the thermal population of the rotational states, a rotational
free energy contribution ($F_{\textrm{rot}}$) must be considered,

\begin{equation}
F_{\textrm{rot}}=-k_{\textrm{B}}T\log(Z_{\textrm{rot}}),\label{eq:free_rot}
\end{equation}
where $Z_{\textrm{rot}}=Z_{\textrm{o}}^{g_{\textrm{o}}}Z_{\textrm{p}}^{g_{\textrm{p}}}$
is the rotational partition function obtained as a factorization of
independent partition functions of ortho ($Z_{\textrm{o}}$) and para
($Z_{\textrm{p}}$) H$_{2}$ molecules, with respective naturally
occurring fractions, $g_{\textrm{o}}=3/4$ and $g_{\textrm{p}}=1/4$
reflect the degeneracy of each spin-isomer \citep{Estreicher2001,Lavrov2002,Estreicher2004,Colonna2012}
, and

\begin{equation}
Z_{\{\textrm{o,p}\}}=\sum_{j=\{\textrm{odd,even}\}}(2j+1)\exp\left[-j(j+1)\theta_{\textrm{rot}}/T\right].\label{eq:z_rot}
\end{equation}
In Eq.~\ref{eq:z_rot} the index $j$ either runs over odd or even
integers for ortho- or para-H$_{2}$, respectively, and the characteristic
rotational temperature of interstitial H$_{2}$ in Si is $\theta_{\textrm{rot}}=73.0$~K
\citep{Estreicher2004}.

We return now to the reference free energy $F^{(0)}$ in Eq.~\ref{eq:free_form}.
This is usually broken into contributions from all $n_{i}$ elements
of species $i$ enclosed within a sample volume, $F^{(0)}=\sum_{i}n_{i}\mu_{i}^{(0)}$.
Here $\mu_{i}^{(0)}=\partial F_{i}^{(0)}/\partial n_{i}$ is the chemical
potential of the $i$-th species, obtained from the free energy, $F_{i}^{(0)}$,
of a reservoir usually considered under normal temperature and pressure
conditions. We are only interested in the calculation of quantities
that involve the conservation of chemical species, including electronic
levels, binding energies, and energy barriers, all of which are independent
of our choice for the chemical potentials. Therefore, we simply set
$\mu_{\textrm{Si}}^{(0)}$ to the free energy per atom in crystalline
silicon at $T=0$~K, which includes electronic plus ionic and zero-point
vibrational energies. As for $\mu_{\textrm{H}}^{(0)}$, it is found
from the free energy per H atom in molecular H$_{2}$ located at the
tetrahedral interstitial site of silicon at $T=0$~K, \emph{i.e.},
$F=2\mu_{\textrm{H}}^{(0)}+n_{\textrm{Si}}\mu_{\textrm{Si}}^{(0)}$.
This choice allows us to directly find the potential energy of a H
defect with respect to the molecular state.

Atomic hydrogen in Si is an impurity that can adopt either positive,
neutral, or negative charge states, depending on the Fermi level of
the sample (here referred with respect to the valence band top energy
$E_{\textrm{v}}$). We define a chemical potential for electrons as
$\mu_{\textrm{e}}=E_{\textrm{v}}+E_{\textrm{F}}$, where value of
$E_{\textrm{v}}$ is assumed to be constant and set to the highest
occupied Kohn-Sham state of a 512-atom bulk calculation at $\mathbf{k}=\Gamma$.
Hence, the free energy reference of Eq.~\ref{eq:free_form} is

\begin{equation}
F^{(0)}=\sum_{i}n_{i}\mu_{i}^{(0)}+n_{\textrm{e}}\mu_{\textrm{e}}=\sum_{i}n_{i}\mu_{i}^{(0)}-q(E_{\textrm{v}}+E_{\textrm{F}}),
\end{equation}
where $E_{\textrm{F}}$ is set \emph{a priori} within the range $0\leq E_{\textrm{F}}\leq E_{\textrm{g}}$,
and $n_{\textrm{e}}=-q$ is the number of extra electrons trapped
at the defect with respect to the neutral charge state ($q=0$).

\subsection{Activation barriers and rates}

Within the statistical formulation of Vineyards's transition state
theory \citep{Vineyard1957,Haenggi1990,Kong2006}, the rate of a thermally
activated process involving atomic motion is given by,

\begin{equation}
\nu=\frac{k_{\textrm{B}}T}{h}\frac{Z^{\ddagger}}{Z},
\end{equation}
where $Z^{\ddagger}$ and $Z$ are partition functions for the transition
state and the initial state, respectively. Again, neglecting electronic
excitations, assuming the harmonic approximation and a dilute regime
(no interactions between defects and negligible impact on the Fermi
level), we arrive at the following Arrhenius rate expression,

\begin{equation}
\nu=\nu_{0}\exp\left(-\frac{\Delta E_{\textrm{a}}}{k_{\textrm{B}}T}\right),\label{eq:nu-arrhenius}
\end{equation}
with the activation energy $\Delta E_{\textrm{a}}=(E_{\textrm{elec}}^{\ddagger}+E_{\textrm{ZP}}^{\ddagger})-(E_{\textrm{elec}}+E_{\textrm{ZP}})$
being obtained from a NEB calculation and the vibrational frequencies
found as described in Sec.~\ref{subsec:ei-energies}. The attempt
frequency is given by

\begin{equation}
\nu_{0}=\frac{Z_{\textrm{rot}}^{\ddagger}}{Z_{\textrm{rot}}}\frac{\prod_{i=1}^{3N-3}\nu_{i}}{\prod_{i=1}^{3N-4}\nu_{i}},\label{eq:prefactor}
\end{equation}
which accounts for differences in rotational and vibrational degrees
of freedom in the initial and transition states. For the migration
of H$_{2}$ molecules we will consider both $Z_{\textrm{rot}}^{\ddagger}$=1
and $Z_{\textrm{rot}}^{\ddagger}=Z_{\textrm{rot}}$, representing
partition functions for static and dynamic (rigid rotor) molecules
at the transition state, respectively. The second fraction on the
right hand side of Eq.~\ref{eq:prefactor} accounts for the phonon
partition functions from the initial (numerator) and transition (denominator)
states, respectively, to the attempt frequency. It is noted that the
transition state only contributes with $3N-4$ vibrational modes,
thus excluding the unstable mode along the jump trajectory.

Atomistic modeling of diffusing impurities in a $d$-dimensional medium
usually starts with Einstein's relation for the diffusivity $D=\langle\Delta r^{2}(t)\rangle/2d\Delta t$,
where $\langle\Delta r^{2}(t)\rangle$ is the mean square impurity
displacement extrapolated to an interval of time $\Delta t\rightarrow\infty$.
If the impurity travels according to random-walk statistics in a three-dimensional
crystal, then we have

\begin{equation}
D=\frac{g_{\textrm{j}}l_{\textrm{j}}^{2}\nu}{6},\label{eq:diffusivity}
\end{equation}
where $\nu$ is the average rate of independent \emph{jumps} of length
$l_{\textrm{j}}$ which can be performed along $g_{\textrm{j}}$ equivalent
paths for each initial state. For instance, bond-centered H has a
total of $g_{\textrm{j}}=6$ available paths to perform a jump to
its neighboring $BC$ sites. Combining Eqs.~\ref{eq:diffusivity}
and \ref{eq:nu-arrhenius} we can write the diffusivity in the Arrhenius
form,

\begin{equation}
D=D_{0}\exp\left(-\frac{\Delta E_{\textrm{a}}}{k_{\textrm{B}}T}\right);\quad D_{0}=\frac{g_{\textrm{j}}l_{\textrm{j}}^{2}\nu_{0}}{6}.\label{eq:arrhenius}
\end{equation}

It is noted that quantum tunneling is not addressed. Such effects
are expected to be more relevant at low temperatures ($T\lesssim80\,K$)
\citep{Herrero1997}.

\section{Results\label{sec:results}}

\subsection{Relative stability of hydrogen monomers and dimers at T=0\label{subsec:stability_zero_t}}

\begin{figure}
\includegraphics{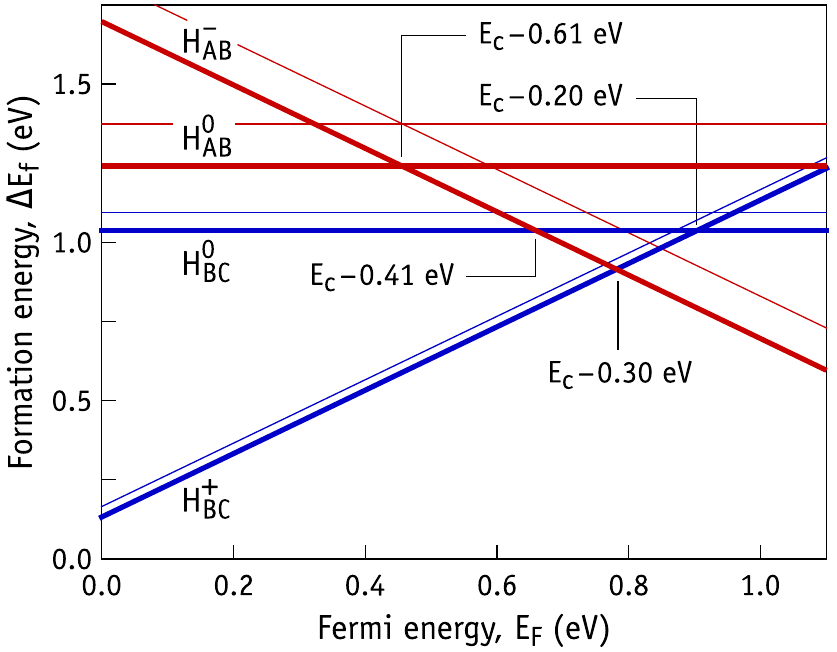}

\caption{\label{fig3}Energy of formation ($\Delta E_{\textrm{f}}$, thick
lines) of hydrogen monomers in Si at $T=0$~K as a function of the
Fermi level ($E_{\textrm{F}}$). Only the electronic and zero-point
vibrational energies are considered. For comparison, we also show
the formation energy as obtained without the contribution of zero-point
motion (thin lines). Calculated transition levels are also indicated.}
\end{figure}

We start by reporting the energetics of hydrogen defects at $T=0$.
In this case Eq.~\ref{eq:free_form} boils down to $\Delta F_{\textrm{f}}(T=0)=\Delta E_{\textrm{f}}$,
essentially involving the computation of the electronic plus ionic
and zero-point vibrational energies of a defective supercell, subtracted
by atomic and electronic chemical potentials,
\begin{equation}
\Delta E_{\textrm{f}}=E_{\textrm{elec}}+E_{\textrm{ZP}}-\sum_{i}n_{i}\mu_{i}^{(0)}+q(E_{\textrm{v}}+E_{\textrm{F}}),
\end{equation}
with $\mu_{i}^{(0)}$ equally accounting for electronic and zero-point
motion of the reference elements. Figure~\ref{fig3} shows the calculated
$\Delta E_{\textrm{f}}$ for several H monomers. We recall that the
chemical potential of H was found from the energy of a molecule at
the $T$ site. For that reason its formation energy is zero (with
and without accounting for zero-point motion). The main features of
the diagram are well known, including the negative-$U$ ordering of
donor and acceptor levels \citep{Walle1988,Walle1989,Herring2001}.
Here we simply improve the results of previous calculations by applying
a non-local functional to the many-electron energy. In practice, the
current results are free from the underestimated band gap \emph{syndrome}
typical of local and semi-local approximations to the exchange-correlation
interactions. It should be mentioned that the \emph{scissors} correction
applied to the local density results of Van de Walle \emph{et~al.}
(Fig.~11 of Ref.~\citep{Walle1989}) turned out to provide a rather
accurate picture. In the present case the calculated band gap, as
found from the energy difference between the lowest unoccupied and
highest occupied Kohn-Sham states, is $E_{\textrm{g}}=1.1$~eV.

\noindent 

Figure~\ref{fig3} shows formation energies that consider the electronic
contributions only (thin lines), along with those that also account
for the zero-point motion as well (thick lines). Blue and red lines
represent $\Delta E_{\textrm{f}}$ values for bond-centered and anti-bonding
configurations, respectively.

In agreement with many previous reports \citep{Walle1988,Walle1989,Herring2001,Estreicher2012},
positively charged H finds its ground state at the bond-center site.
Also like in other reports, the neutral state has two close energy
structures, namely the ground state at the $BC$ site, and a metastable
state (0.20~eV above) with H located within the tetrahedral interstitial
cage.

We find that the energy of tetrahedral interstitial H$_{T}^{0}$ and
H$_{T}^{-}$ states are higher than H$_{AB}^{0}$ and H$_{AB}^{-}$
by respectively 0.15 and 0.04~eV. Although such small figures have
little impact when estimating the location of the transition levels,
it is important to refer that the H$_{T}^{0}$ state in particular
is definitely a local maximum of energy, whose height is comparable
to the barrier for reorientation of H$_{AB}^{0}$ within the $T$
cage. Full HSE-level atomistic relaxations using smaller 64-Si-atom
supercells confirm these findings within less than 10~meV. Another
indication that H$_{T}^{0}$ is not stable comes from a triply degenerate
set of negative eigenvalues of the dynamical matrix for this structure,
corresponding to an imaginary frequency of $\nu=\textrm{i}\,520~\textrm{cm}^{-1}$
(in addition to three translational modes). As for the negative state,
the potential is very flat. Considering zero-point energy and zero-point
motion effects, the energy difference between tetrahedral and anti-bonding
structures (40~meV) becomes irrelevant, and most probably H$^{-}$
roams freely within the whole volume of the $T$ cage even at $T\sim0$~K.
The lower energy H$_{AB}^{-}$ was the structure employed in the calculations
presented below (unless specified otherwise). It is unclear why H$_{T}$
is a maximum in the potential energy surface. The defect introduces
a singlet state in the gap, and we can only suggest that the off-site
location could be driven by a pseudo-Jahn-Teller effect.

Regarding the calculated charge state transition levels, we obtain
a donor transition at $E_{\textrm{c}}-0.20$~eV involving the H$_{BC}$
structure (blue lines in Figure.~\ref{fig3}). This is very close
to the $E3'$ electron trap measured at $E_{\textrm{c}}-0.175$~eV
\citep{Irmscher1984,Nielsen1999}. An acceptor at $E_{\textrm{c}}-0.61$~eV
($E_{\textrm{v}}+0.49$~eV) was also found, but now involving the
H$_{AB}$ configuration (red lines in Figure.~\ref{fig3}). Considering
that H$_{BC}^{0}$ is the actual ground state for neutral H (lying
0.2~eV below H$_{AB}^{0}$), the thermodynamic acceptor level is
therefore at $E_{\textrm{c}}-0.41$~eV ($E_{\textrm{v}}+0.69$~eV),
which combined with the donor transition leads to a negative-$U$
$(-/+)$ level at $E_{\textrm{c}}-0.30$~eV. Both the acceptor and
negative-$U$ levels edge the ranges $E_{\textrm{c}}-E(-/0)=0.4\textrm{-}0.6$~eV
and $E_{\textrm{c}}-E(-/+)=0.3\textrm{-}0.4$~eV, estimated by junction
capacitance measurements \citep{Nielsen2002}.

According to Van de Walle and Neugebauer \citep{Walle2003}, a universal
charge neutrality level (UCNL) that pins the $(-/+)$ transition of
hydrogen was postulated at about $-4.5$~eV with respect to vacuum.
For the case of Si, this is located at $E_{\textrm{c}}-4.5\:\textrm{eV}+\chi=E_{\textrm{c}}-0.4$~eV,
where $\chi=4.1$~eV is the electron affinity of crystalline Si \citep{Fujimura2016}.
Importantly, if we allow for (i) some spread in the $(-/+)$ transition
estimated by the UCNL method (due to for instance dielectric screening
variance — a spread of about 0.4~eV was found after considering a
range of semiconductors and insulators \citep{Walle2003}), (ii) for
the uncertainty in the measurements (\emph{c.a.} 0.1~eV) and (iii)
for the typical error bar of our calculations (\emph{c.a.} 0.1~eV),
we find an overlap and consistency among all three methods.

Figure~\ref{fig3} shows that zero-point motion effects can have
a sizable impact regarding the location of the levels. Bond-centered
hydrogen defects have a zero-point vibrational contribution $\Delta E_{\textrm{ZP}}=E_{\textrm{ZP}}-E_{\textrm{ZP}}^{(0)}\sim0.1$~eV,
where $E_{\textrm{ZP}}^{(0)}$ is the zero-point energy of the reference
elements (bulk Si and molecular hydrogen in Si). This figure is about
twice as much found for anti-bonding monomers (see Table~\ref{tab1}).
The formation energies incorporating the zero-point motion (thick
lines in Figure~\ref{fig3}), are below the electronic formation
energies (thin lines) because their respective references differ by
$\Delta E_{\textrm{ZP}}(\textrm{H}_{2T})/2=0.18$~eV, which is the
zero-point motion contribution to the H chemical potential. The results
are qualitatively in line with those obtained by Karazhanov and co-workers
\citep{Karazhanov2014}, but quantitatively different. We find deviations
of just above $\sim0.1$~eV in the location of the electronic levels
after considering zero-point energies, perhaps because we considered
the vibrational modes of all atoms in the supercells. In Ref.~\citep{Karazhanov2014}
only a few atoms around the defects were considered for constructing
the dynamical matrix.

As for the relative stability of hydrogen dimers we find that H$_{2}^{*}$
is metastable by 0.19~eV with respect to H$_{2T}$. This result includes
the zero-point energy difference (0.01~eV higher in H$_{2}^{*}$)
and it is in line with previous reports as well (see Ref.~\citep{Estreicher2014}
and references therein). Based on such a small energy difference,
it is difficult to understand the exclusive formation of molecular
H$_{2}$ in samples quenched from high temperatures.

We could not find any additional dimer whose stability was comparable
to that of H$_{2T}$ and H$_{2}^{*}$. The third most favorable geometry
for a pair of hydrogen atoms in silicon, was made of two anti-bonding
Si-H$_{AB\ensuremath{}}$ units sitting on opposite sides of a Si-Si
broken bond. This structure, previously referred to as H$_{2}^{**}$
\citep{Hourahine2000}, was less stable than H$_{2T}$ by 1.1~eV.
Hence, our results cannot explain the existence of a third dimer (H$_{2\textrm{B}}$)
as proposed in Ref.~\citep{Voronkov2017a}, unless its formation
involves an extrinsic impurity that provides a stabilizing effect.
However, this possibility presents difficulties regarding an explanation
for the fast diffusing nature of H$_{2\textrm{B}}$, whose activation
energy was estimated to be as low as 0.53~eV \citep{Voronkov2017a}.

\subsection{Migration and reconfiguration\label{subsec:migration}}

\begin{figure*}
\includegraphics{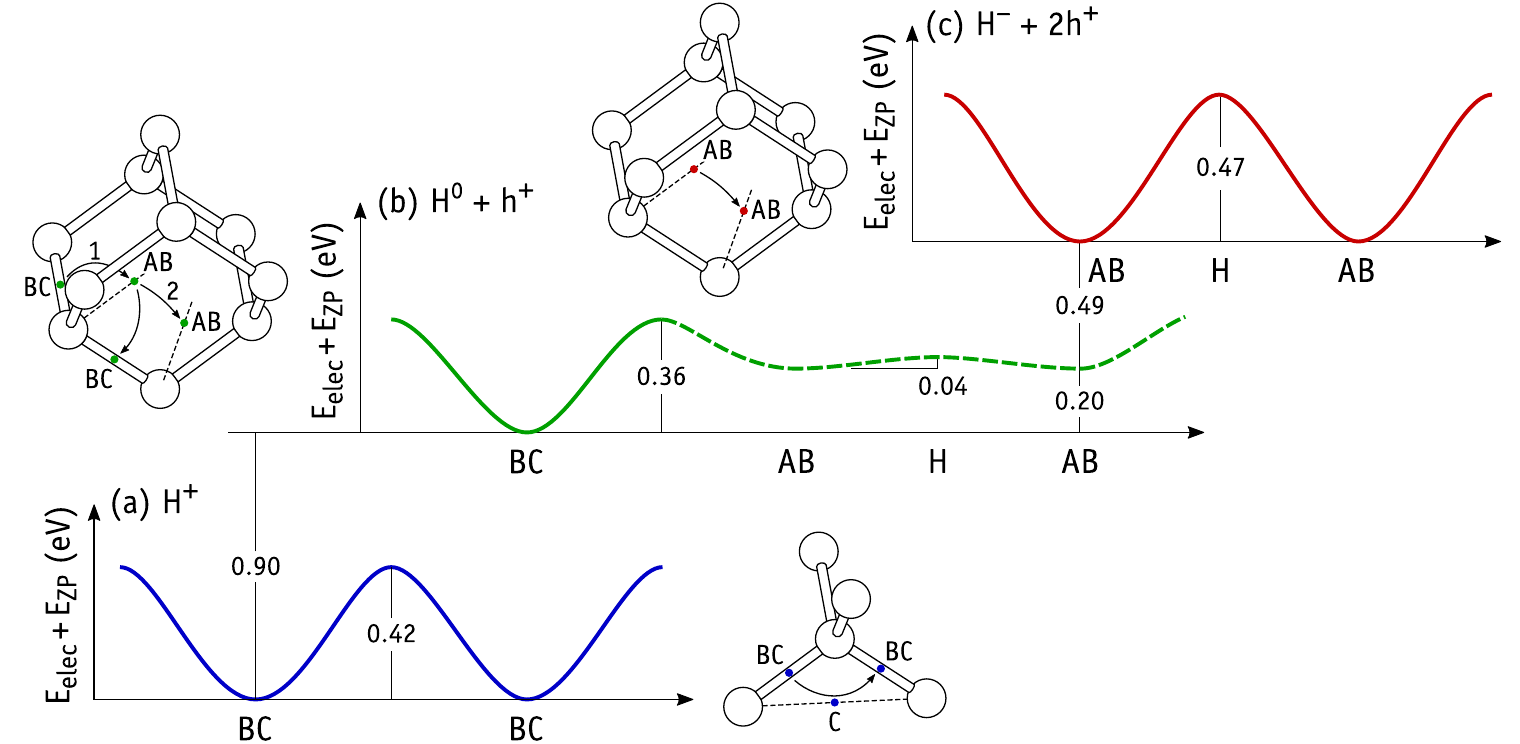}

\caption{\label{fig4}Calculated configuration coordinate diagram for atomic
hydrogen in silicon. All minima and barriers indicated account for
electronic (plus ionic) and zero-point vibrational energies. CCD for
charge states plus (a), neutral (b) and negative (c) are offset in
the energy scale by $E(0/+)-E_{\textrm{v}}=0.90$~eV and $E(-/0)-E_{\textrm{v}}=0.69$~eV,
corresponding to the energy for hole emission from H$_{BC}^{+}$ and
H$_{BC}^{0}$ ground states to the valence band top.}
\end{figure*}

Figure~\ref{fig4} shows a configuration coordinate diagram (CCD)
for atomic hydrogen in p-type Si, constructed from the formation energies
at $T=0$ and calculated migration/transformation barriers also at
$T=0$. The transition states were found using a 13-image-NEB method
(see Sec.~\ref{subsec:ei-energies}) and account for the zero-point
energy of $3N-4$ vibrational modes. The three charge states, namely
positive (blue line), neutral (green line) and negative (red line)
are offset in the energy scale by hole emission energies $E(0/+)-E_{\textrm{v}}=0.90$~eV
and $E(-/0)-E_{\textrm{v}}=0.69$~eV.

\noindent 

As expected, the migration mechanism of H$_{BC}^{+}$ was found along
the shortest path passing close to the $C$-point. The process is
depicted in the structure of Fig.~\ref{fig4}(a). The transition
state lies 0.42~eV above the ground state, underestimating the value
measured at low-temperatures merely by 0.01~eV \citep{Gorelkinskii1996},
and underestimating the high-temperature data by 0.06~eV \citep{vanWieringen1956}.

An interesting property of the anti-bonding configuration of H$_{AB}^{-}$
is that the migration mechanism is not along the $\langle111\rangle$
crystalline directions (passing by the hexagonal site) as previously
assumed for H$_{T}^{-}$. Instead the hydrogen moves through the hexagonal
channels along $\langle110\rangle$ directions as depicted in Fig.~\ref{fig4}(c).
Still the transition state is very close to the hexagonal site and
the estimated barrier is $\Delta E_{\textrm{a}}=0.47$~eV, i.e.,
0.05~eV higher than that for the migration of H$_{BC}^{+}$. Consistently
with previous theoretical estimates, the barrier is much lower than
the 0.7~eV activation energy obtained from the kinetic studies of
the release/re-trapping of H by dopants upon injection of minority
carriers \citep{Johnson1992}.

Now we look at the neutral state. Hole emission from the H$_{BC}^{0}$
ground state must be preceded by a reconfiguration to the $\textrm{H}{}_{AB}^{0}$
metastable state. This process is represented by the arrow with label
``1'' in the structure of Figure~\ref{fig4}(b). The potential
energy close to $\textrm{H}{}_{AB}^{0}$ is represented in the CCD
as a dashed green line with a minimum 0.2~eV above H$_{BC}^{0}$.
The calculated barrier for the $\textrm{H}{}_{BC}^{0}\rightarrow\textrm{H}{}_{AB}^{0}$
reconfiguration is $\Delta E_{\textrm{a}}=0.36$~eV. However, since
the barrier for returning to the $\textrm{H}{}_{BC}^{0}$ ground state
is only $\Delta E_{\textrm{a}}=0.16$~eV, hole emission from the
metastable state $\textrm{H}{}_{AB}^{0}\rightarrow\textrm{H}{}_{AB}^{-}+h^{+}$
(with emission energy $\Delta E=0.49$~eV) is highly unlikely. This
makes the direct observation of the $(-/0)$ transition an extremely
difficult task to be performed by charge-countable techniques such
as DLTS.

In line with previous results \citep{Estreicher2014}, the calculated
barrier for the hopping of a neutral H atom between $BC$ sites is
considerably higher than that between $AB$ sites. Small differences
were found though. We find that the minimum energy path between two
neighboring $\textrm{H}{}_{BC}^{0}$ states is along the $\textrm{H}{}_{AB}^{0}$
configuration, corresponding to the already mentioned mechanism shown
by the arrow ``1'' in Figure~\ref{fig4}(b), with a barrier of
$\Delta E_{\textrm{a}}=0.36$~eV. Jumps in the direction of the $C$
point {[}\emph{c.f.} structure of Fig.~\ref{fig4}(a){]} correspond
to a barrier $\Delta E_{\textrm{a}}=0.44$~eV high. On the other
hand, jumps between $\textrm{H}{}_{AB}^{0}$ states along the hexagonal
channel in the $\langle110\rangle$ direction (arrow ``2'' in Figure~\ref{fig4}(b))
involve surmounting a minute barrier of only 0.04~eV (including zero-point
energy contributions).

Of course, depending on the temperature, doping type and concentration,
the presence of photogenerated or injected minority carriers, the
hydrogen atom can reside temporarily in the $\textrm{H}{}_{AB}^{0}$
state and perform several jumps along such a flat potential landscape
before returning to the H$_{BC}^{0}$ ground state. Essentially, the
migration of neutral H could be described by a thermally activated
$\textrm{H}{}_{BC}^{0}\rightarrow\textrm{H}{}_{AB}^{0}$ jump ($\Delta E_{\textrm{a}}=0.36$~eV followed by the reverse process)
with an effectively long traveling distance that depends on the life-time
of the metastable $\textrm{H}{}_{AB}^{0}$ state. This effect is qualitatively
addressed in Sec.~\ref{subsec:diffusivity}.

The lowest energy configuration of the hydrogen molecule in Si has
the H-H bond along $\langle111\rangle$ directions. In agreement with
previous studies, we found that H$_{2T}$ can rotate almost freely
in the $T$ site \citep{Estreicher2001,Estreicher2004}. The energy
of other molecular alignments differed by less than 5~meV.

We also investigated the reconfiguration and migration of hydrogen
dimers. Regarding the transformation $\textrm{H}_{2T}\rightarrow\textrm{H}_{2}^{*}$
we found that the most favorable mechanism involves the collision
of $\textrm{H}_{2T}$ molecules with Si-Si bonds. Taking for instance
a molecule at the central $T$ site of Figure~\ref{fig1}, the collision
with one of the 12 nearest $BC$ sites (\emph{e.g.} the one indicated
with a blue dot) results in the breaking of the H-H bond and the formation
of a metastable $(\textrm{H}_{BC}\textrm{-}\textrm{H}_{AB})^{*}$
complex. This is expected to be the rate-limiting step for the breaking
of H$_{2T}$ without the assistance of defects. In this complex the
Si-H$_{BC}$-Si unit is slightly bent toward the $C$ site, while
the $\textrm{H}_{AB}$ atom is very close to the original $T$ site
of the molecule. The reaction $\textrm{H}_{2T}\rightarrow(\textrm{H}_{BC}\textrm{-}\textrm{H}_{AB})^{*}$
has an activation energy of $\Delta E_{\textrm{a}}=1.62$~eV and
the resulting metastable complex is $\Delta E=1.14$~eV higher in
energy than $\textrm{H}_{2T}$. Further displacement of the $\textrm{H}_{BC}$
unit to finally form H$_{2}^{*}$ involves surmounting a barrier of
only 0.40~eV, \emph{i.e.}, the second transition state is 1.54~eV
above the initial $\textrm{H}_{\textrm{2T}}$ state.

Instead of breaking, the $\textrm{H}_{2T}$ molecule can migrate across
the lattice. The migration of H$_{2T}$ in Si is commonly described
as involving the motion of the molecules between $T$ sites through
$H$ sites with their H-H bonds parallel to the direction of motion
(perpendicular to the hexagonal rings). Our results suggest that this
is not the best picture. We found that the energy of the molecule
at the $H$ site depends very little on its crystalline orientation.
In fact, the transition state along the minimum energy path is the
one with the H-H bond perpendicular to the direction of motion. Considering
zero-point motion effects, the barrier is $\Delta E_{\textrm{a}}=0.82$~eV
high, and this is only 0.02~eV lower than if we considered a transition
state with the H-H bond along the $\langle111\rangle$ direction of
motion. This figure is very close to the activation energy $\Delta E_{\textrm{a}}=0.78$~eV
for the capture kinetics of migrating H$_{2}$ molecules by interstitial
oxygen impurities \citep{Markevich1998}. In Section~\ref{subsec:diffusivity}
we will argue that the molecules should be treated as nearly free-rotors
all along the migration path, including at the transition state.

The above results demonstrate that the barrier for breaking the molecules
is almost twice that for the migration. Assuming a typical pre-exponential
factor of $\nu_{0}\sim10^{12}$~s$^{-1}$ for a thermally activated
process, a dissociation rate of $\nu\sim1$~s$^{-1}$ for the breaking
of the molecules is reached when the temperature is nearly 400~$^{\circ}$C.
It appears that the lifetime of H$_{2T}$ significantly exceeds that
of H$_{2}^{*}$ (according to the experimental results, H$_{2}^{*}$
is not thermally stable above 200~$^{\circ}$C \citep{Holbech1993}).
Therefore, it can be suggested that during quenching from high temperatures,
there is a wide temperature window of $\sim200\textrm{-}400$~$^{\circ}$C
where the molecule can form, migrate and react with defects and impurities,
avoiding formation of H$_{2}^{*}$.

\subsection{Stability of hydrogen species at finite temperatures\label{subsec:stability_finite_t}}

\begin{figure}
\includegraphics{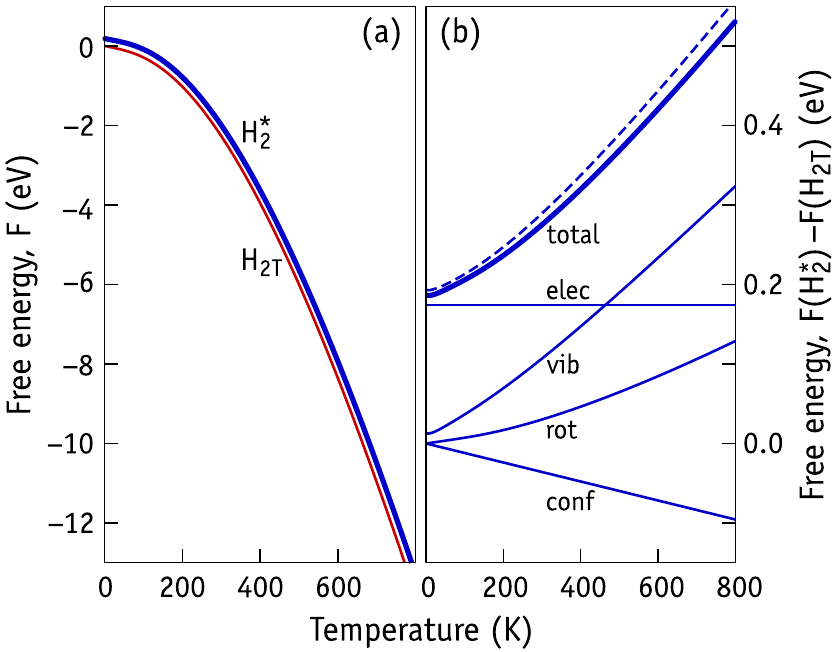}

\caption{\label{fig5}(a) Free energy of H$_{2}^{*}$ (blue line) and H$_{2T}$
(red line) in the temperature range $T=0\textrm{-}800$~K. (b) Difference
between free energies of H$_{2}^{*}$ and H$_{2T}$ in silicon, $F(\textrm{H}_{2}^{*})-F(\textrm{H}_{2T})$,
represented as a thick solid line. Partial contributions from the
electronic plus ionic potential (elec), vibrational modes calculated
in 64 Si atom cells (vib), rotational states of H$_{2T}$ (rot) and
configurational entropy (conf), are also show as thin solid lines.
The dashed curve represents the same quantity obtained using a larger
supercell with 216 atoms for the evaluation of the vibrational free
energy.}
\end{figure}

We start the reporting of the finite temperature results with a comparison
of the relative stability of H$_{2T}$ and H$_{2}^{*}$. The results
are summarized in Figure~\ref{fig5}, where we show the total free
energy of both defects calculated according to Eq.~\ref{eq:free_tot}
in the temperature range $T=0\textrm{-}800$~K on the left, and 
their difference, $\Delta F_{\textrm{dimers}}=F(\textrm{H}_{2}^{*})-F(\textrm{H}_{2T})$,
as a thick solid line on the right. The thin solid lines in Figure~\ref{fig5}(b)
represent partial contributions from the electronic plus ionic potential
(elec), vibrational modes calculated in 64 Si atom cells (vib), rotational
states of H$_{2T}$ (rot) and configurational entropy (conf). For
this specific case, the configurational entropy accounts for the different
number of orientations of each species and does not dependent on the
concentration of hydrogen in the silicon.

The dashed line in Figure~\ref{fig5}(b) represents the formation
energy difference between H$_{2}^{*}$ and H$_{2T}$, with the vibrational
free energy obtained from larger 216-Si-atom cells (645 normal modes
of vibration). Clearly, the number of phonons represented by the smaller
cells are sufficient for the present purpose.

\noindent 

At a glance, Figure~\ref{fig5}(a) does not provide us with much
information, apart from the fact that no matter the temperature, H$_{2}^{*}$
is less stable than H$_{2T}$. More interesting conclusions can be
drawn from analysis of $\Delta F_{\textrm{dimers}}$ shown in Figure~\ref{fig5}(b).
In this figure it is clear that $\textrm{H}_{2}^{*}$ becomes less
favorable with raising the temperature, or alternatively, the molecule
becomes progressively more stable. These findings are in line with
previous results by Estreicher and co-workers \citep{Estreicher2004}.

The electronic component of $\Delta F_{\textrm{dimers}}$, evaluated
from 512 atom cells plus defects, gives $\Delta E_{\textrm{elec}}=0.17$~eV.
Zero-point motion raises $\Delta F_{\textrm{dimers}}(T=0)$ to $\Delta E=0.19$~eV.
At finite temperatures, the dominant contribution to $\Delta F_{\textrm{dimers}}$
comes from the vibrational degrees of freedom. The effect of molecular
rotation cannot be neglected and nearly compensates for the configurational
entropy difference between both defects (the number of possible locations/orientations
per unit cell is $g=8$ and 2 for $\textrm{H}_{2}^{*}$ and $\textrm{H}_{2T}$,
respectively). At $T=500$~K, which is approximately the annealing
temperature of $\textrm{H}_{2}^{*}$, the free energy difference increases
to $\Delta F_{\textrm{dimers}}(T=500\:\textrm{K})=0.37$~eV. At $T\sim700$~K,
which is our estimated temperature for the breaking of the molecules
due to collisions against the Si-Si bonds, $\Delta F_{\textrm{dimers}}$
raises to almost 0.5~eV. Again, this effect is consistent with the
observed formation of molecules (and not $\textrm{H}_{2}^{*}$ dimers)
when the Si is quenched after heat-treatments at temperatures higher
700~$^{\circ}$C in the presence of a hydrogen source.

\begin{figure}
\includegraphics{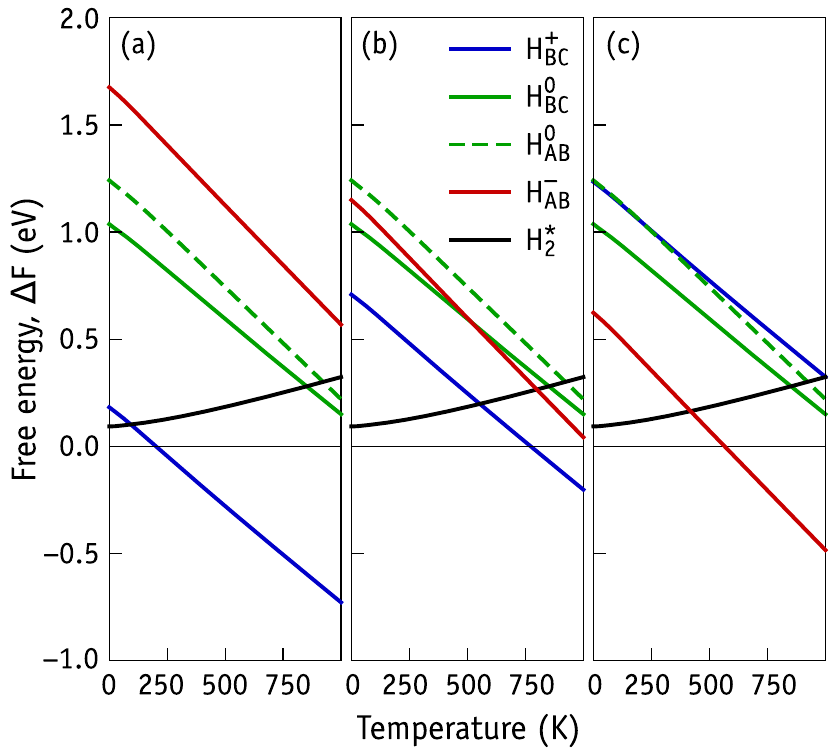}

\caption{\label{fig6}(a) Temperature dependence of the free energy of formation
per hydrogen atom for several H-defects in Si with respect to the
same quantity for molecular H$_{2}$ in Si. (a) For the Fermi level
at the valence band top, (b) for the Fermi level at mid-gap, (c) for
the Fermi level at the conduction band bottom. The configurational
entropy was evaluated considering a concentration of H monomers and
dimers of $2\times10^{14}$~cm$^{-3}$ and $1\times10^{14}$~cm$^{-3}$,
respectively.}
\end{figure}

Figure~\ref{fig6} depicts the free energy change per H atom, $\Delta F=F(X_{n})/n-F(\textrm{H}_{2T})/2$,
of several H defects $X=\{\textrm{H}_{2}^{*},\,\textrm{H}_{BC}^{+},\,\textrm{H}_{BC}^{0},\,\textrm{H}_{AB}^{0},\,\textrm{H}_{AB}^{-}\}$
with $n=\{1,2\}$ being the number of H elements in each defect. The
zero energy refers to a sample of bulk silicon with a concentration
$1\times10{}^{14}$~cm$^{-3}$ of H$_{2T}$ molecules at a specific
temperature $T$. The calculations were carried out in the temperature
range $T=0\textrm{-}1000$~K. For the calculation of the configurational
entropy, concentrations of $2\times10^{14}$~cm$^{-3}$ and $1\times10^{14}$~cm$^{-3}$
were assumed for H monomers and dimers, respectively. Three distinct
situations are considered in the figure, namely when the Fermi energy
is located (a) at the top of the valence band; (b) at mid-gap; (c)
at the bottom of the conduction band. Figure~\ref{fig6}(b) is the
one that intends to reproduce the intrinsic conditions attained at
high temperatures.

\noindent 

Regarding the $\textrm{H}_{2}^{*}$ (black line) the plot essentially
reproduces the results of Figure~\ref{fig5}(b). As for the monomers,
their free energy with respect to that of H$_{2T}$ decreases with
$T$. This is mostly a configurational entropy effect which notably
makes H$_{BC}^{+}$ and H$_{AB}^{-}$ more stable than H$_{2T}$ in
p-type and n-type Si above $T\sim250$~K and 550~K, respectively
(\emph{c.f.} Figs.~\ref{fig6}(a) and \ref{fig6}(c)). However, at
these temperatures and above, any existing molecule that was not trapped
by other impurities or defects, must overcome the 1.6~eV dissociation
barrier before the hydrogen atoms can capture holes or electrons to
end up as 2H$_{BC}^{+}$ or 2H$_{AB}^{-}$, respectively. Of course,
H$_{BC}^{+}$ and H$_{AB}^{-}$ are fast diffusing species and anneal
out well below room temperature.

Under intrinsic conditions and at temperatures high enough to release
the hydrogen from traps, H$_{BC}^{+}$ becomes the most favorable
hydrogen defect above $T\sim750$~K ($T\sim480$~$^{\circ}$C).
This provides an estimate for the temperature below which H$_{2T}$
formation becomes thermodynamically favorable with respect to the
dissociated state.

The rate of change of the free energies with the temperature are also
different among the H species. The free energy of bond-centered defects,
shown in Figure~\ref{fig6} as solid blue and solid green lines,
have a similar (almost parallel) change with $T$, dominated by the
configurational entropy. The vibrational and rotational (from the
molecule reference) free energy contributions have opposite signs
(negative and positive, respectively) and almost cancel. The decreasing
rate of $\Delta F$ for the anti-bonding configurations is slightly
faster, especially for the negatively charged H$_{AB}^{-}$ state
(solid red line). We will further investigate this effect in the future.
Presently, we can only suggest that the anti-bonding configurations
soften the neighboring Si-Si bonds. That leads to a \emph{compression}
of the spacing between the excited vibrational states of crystalline
modes localized around the defect, to an increasingly number of accessible
states at finite temperatures, and therefore, to an increase of the
vibrational entropy, $-TS_{\textrm{vib}}$.

Interestingly, because of the above effect, at $T\gtrsim500$~K under
intrinsic conditions H$_{AB}^{-}$ becomes the second most stable
monomer after H$_{BC}^{+}$. These result allows us to propose the
following interpretation of the effective formation of molecular hydrogen
during the cooling of hydrogenated samples from high temperatures
($T\gtrsim700$~$^{\circ}$C): (1) above $T\sim700$~K the molecules
are not stable and free hydrogen may only be detected transiently
in the atomic form, mostly as H$_{BC}^{+}$, but also as a small fraction
of H$_{AB}^{-}$; (2) Below $T\sim700$~K the interaction between
H$_{BC}^{+}$ and H$_{AB}^{-}$ leads to the formation of stable H$_{2T}$
molecules. (3) With further cooling, H$_{AB}^{-}$ is further consumed
by H$_{\textrm{BC}}^{+}$, and formation of molecules occur as long
as the temperature remains above $T\sim500$~K, below which the fraction
of H$_{AB}^{-}$ becomes negligible in comparison to other states,
and H$_{2T}$ formation stops.

We will also argue in the next section, that the formation of H$_{AB}^{-}$
at high temperatures has implications to the diffusivity of hydrogen.

\subsection{Diffusivity of hydrogen in silicon\label{subsec:diffusivity}}

In this section we report our results pertaining the diffusivity of
hydrogen species, including the effects of temperature. Migration
barriers were already reported in Section~\ref{subsec:migration},
and now we look at the attempt frequencies (Eq.~\ref{eq:prefactor}),
temperature dependent jump rates (Eq.~\ref{eq:nu-arrhenius}) and
diffusivities (Eq.~\ref{eq:arrhenius}).

\noindent 
\begin{figure*}
\includegraphics{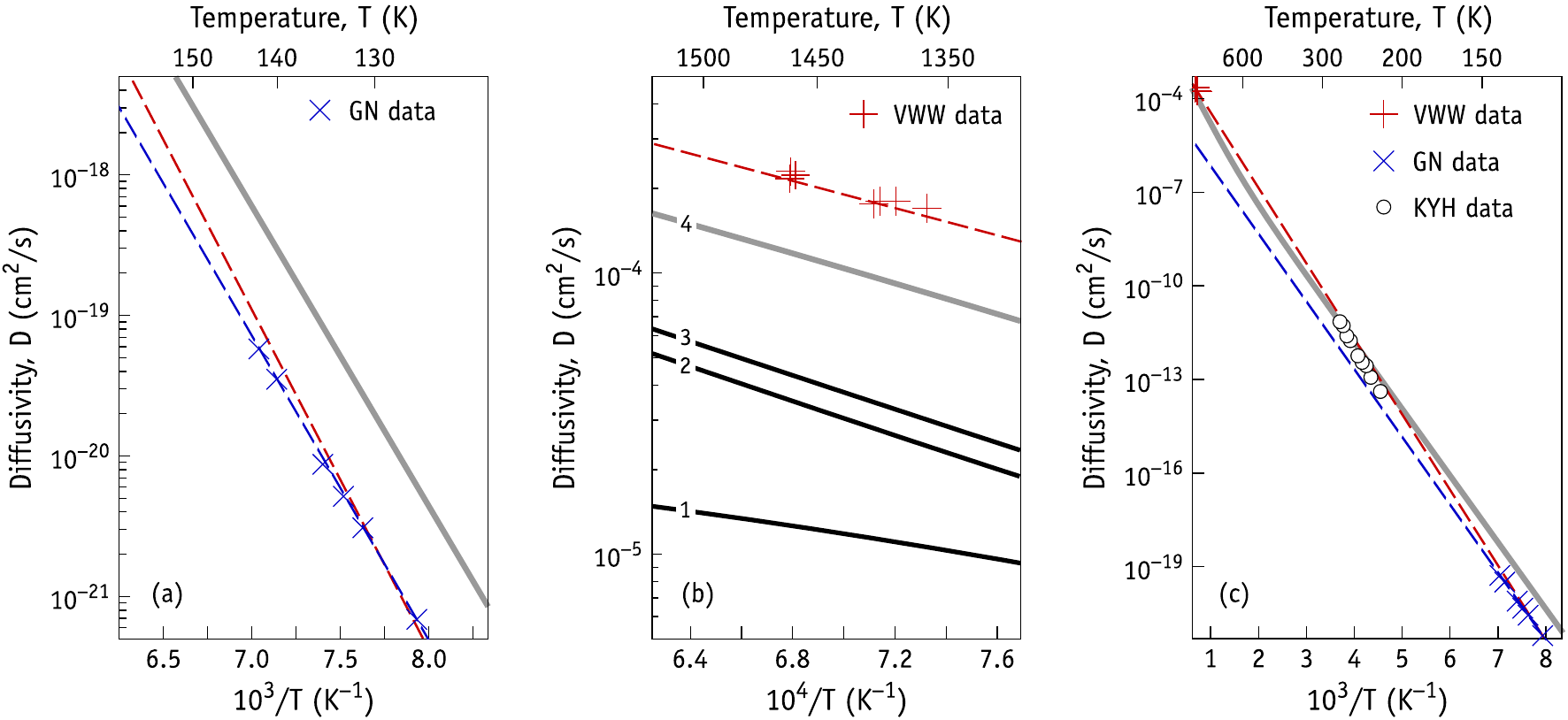}

\caption{\label{fig7}Calculated diffusivity (solid gray line) of atomic hydrogen
in silicon (solid lines) along with measured data (symbols and dashed
lines). (a) Low temperature region with the GN data \citep{Gorelkinskii1996},
extrapolated VWW diffusivity and calculated diffusivity (gray solid
line). (b) High temperature region with the VWW data \citep{vanWieringen1956}
along with the calculations showing the cumulative effect of considering
(1) H$_{BC}^{+}$, (2) H$_{AB}^{-}$, (3) H$_{BC}^{0}$, and (3) H$_{AB}^{0}$
species to the weighted total diffusivity (4). (c) Extrapolation of
the diffusivities from the high temperature and low temperature data
using Arrhenius relations (dashed lines). The intermediate temperature
data of KYH \citep{Kamiura1991} is also shown.}
\end{figure*}

The calculations were carried out considering the Fermi level pinned
to mid-gap. This impacts free energies of formation and the mix of
H monomers participating in atomic hydrogen transport. On the other
hand, given that H$_{2T}$ is electrically inactive, results pertaining
its diffusivity are independent of the doping type.

Below we discuss the diffusivity of H monomers in the context of the
high temperature measurements of VWW ($T\sim1400$~K) and the low
temperature measurements of GN ($T\sim130$~K). At high temperatures,
dopants are expected to play a negligible effect on H diffusivity
and the results are valid for any doping type. At low temperatures,
the results represent the diffusivity of H$^{+}$ alone (or H$^{+}$
elementary jumps as monitored in the GN experiments). They are applicable
to intrinsic and p-type Si, and do not account for eventual trapping
effects by dopants and impurities. Molecular diffusivity can be neglected
at such low temperatures.

Figure~\ref{fig7} shows three Arrhenius plots of the diffusivity
of atomic hydrogen versus temperature in different temperature ranges.
In Fig.~\ref{fig7}(a) the low temperature measurements by Nevinnyi
and Gorelkinskyy are represented by crosses (GN data) and the best
Arrhenius fit to the data gives $\nu_{0}=0.3\:\textrm{THz}$ and $\Delta E_{\textrm{a}}=0.43\,\textrm{eV}$
\citep{Gorelkinskii1996,Herring2001} (blue dashed line). These figures
pertain to the jump of H$_{BC}^{+}$ between bond-center sites because
the stress-alignment of the hydrogen defects in these experiments
was carried out in darkness, where only anisotropic H$_{BC}^{+}$
could form. Considering the $3N-3=189$ phonon frequencies obtained
for the ground state of H$_{BC}^{+}$, as well as the analogous $3N-4=188$
phonon frequencies for the transition state between two neighboring
$BC$ sites, we arrive at a frequency of attempt $\nu_{0}=1.4\:\textrm{THz}$,
nearly a factor of five higher than the figure extracted from the
measurements. Considering that during migration each proton can perform
$g_{\textrm{j}}=6$ equivalent jumps of length $l_{\textrm{j}}=1.92$~Å,
we arrive at a calculated diffusivity $D(T)=(5.3\times10^{-4}\:\textrm{cm}^{2}/\textrm{s}^{-1})\,\exp(-0.42\,\textrm{eV}/k_{\textrm{B}}T)$.
This result is shown in Figure~\ref{fig7}(a) as a solid gray line,
above the experimental diffusivity. Considering the error bars typically
involved in the calculation of $\nu_{0}$ (easily up to a factor of
5), we take the match between the calculations and the measurements
as acceptable.

Comparing the measured low-$T$ diffusivity with the measurements
extrapolated from the VWW data (red dashed line in Fig.~\ref{fig7}(a)),
we find a surprising agreement. Of course the potential errors in
such extreme extrapolation must be considered, particularly as the
VWW data has noticeable scatter and was measured only over a relatively
narrow range of temperatures. Yet if we insist in pursuing such an
exercise, we find that at high temperatures the diffusivity is enhanced
by several dacades (Fig.~\ref{fig7}(c)). The obvious question would
be: What is the route cause for the \emph{bowing} in the hydrogen
diffusivity in Si?

Considering that the path-integral Monte Carlo calculations by Herrero
\citep{Herrero1997} indicate that tunneling motion of bond-centered
H in silicon only takes place below $T\lesssim80$~K, the low-$T$
GN data of Figure~\ref{fig7}(a) is expected to be well described
by classical transition-state theory. It is therefore more likely
that any eventual acceleration in the diffusivity should occur at
high temperatures. The high-$T$ data is shown in Figure~\ref{fig7}(b)
as ``plus'' symbols. The best fit to the data, shown as a red dashed
line, gives $D=(9.4\times10^{-3}\,\textrm{cm}^{2}/\textrm{s})\exp(-0.48\pm0.05\,\textrm{eV}/k_{\textrm{B}}T)$,
and it is above the extrapolated low-temperature diffusivity (blue
dashed line) by a factor of $\sim60$.

It was proposed in Ref.~\citep{Voronkov2017b} that a fast diffusing
neutral hydrogen, although present in small quantities even at $T\sim1000$~$^{\circ}$C,
could lead to the observed enhancement of the diffusivity. Based on
the calculated free energies of all four studied hydrogen monomers
(H$_{BC}^{+}$, H$_{AB}^{-}$, H$_{BC}^{0}$ and H$_{AB}^{0}$), we
estimated their equilibrium fractional concentrations as a function
of temperature as $c_{m}=p_{m}/Z$, with $p_{m}=\exp(-F_{m}/k_{\textrm{b}}T)$,
$Z=\sum_{m}p_{m}$ and $F_{m}$ is the temperature dependent free
energy of species $m$ (which is an index that runs over all monomers).
For the calculation of the free energies, the Fermi level was assumed
to be locked at mid-gap, which is adequate to describe the high temperature
conditions. An effective diffusivity $D_{\textrm{eff}}=\sum_{m}c_{m}D_{m}$
was evaluated from calculated attempt frequencies for jumping of all
four species ($\nu_{0}$), activation energies for migration ($\Delta E_{\textrm{a}}$),
number of equivalent jumps ($g_{\textrm{j}}$) and respective lengths
($l_{\textrm{j}}$). All these data are summarized in Table~\ref{tab1}.

Figure~\ref{fig7}(b) shows the cumulative effect of progressively
adding the contribution of migrating species to the total diffusivity.
The line with label 1 represents the diffusivity obtained when only
H$_{BC}^{+}$ is considered. This is and extrapolation of the Arrhenius relation represented
by the thick gray line in Figure\ref{fig7}(a). At 1500~K the calculated
H$_{BC}^{+}$ diffusivity it is only $\sim2.5$ times higher than the
value extrapolated from the low-$T$ measurements (blue dashed line),
but about 20 times lower than the high-$T$ data. At the same temperature,
the diffusivity is enhanced by $\sim3$ times if in addition to H$_{BC}^{+}$
we consider the presence of negatively charged H$_{\textrm{AB}}^{-}$
species (line 2). Under these conditions, H$_{AB}^{-}$ is the second
most stable monomer (see Fig.~\ref{fig6}(b)) with a free energy
of 0.12~eV above H$_{\textrm{BC}}^{+}$ and a relative population
$p\approx1/4$. Although the calculated activation energy for migration
($\Delta E_{\textrm{a}}=0.47$~eV) is slightly higher than that of
the proton, the pre-exponential factor is also larger, almost by a
factor of 7. A large pre-exponential factor indicates a saddle-point
that is ``wide'' in the configurational space and ``highly probable
to surmount'' , as opposed to a small frequency of attempt which
is typical of ``narrow'' transition states which are ``less probable
to surmount''.

If we continue with adding the remaining monomers to the diffusing
mix, we find that the most stable bond-centered neutral hydrogen leads
to a small enhancement (line 3), mostly due to its minute concentration.
However, the free energy of H$_{AB}^{0}$ decreases faster with temperature
than that H$_{BC}^{0}$ (see Figure~\ref{fig6}(b)), and at $T=1500$~K
(intrinsic Si) both states essentially become degenerate, with a free
energy about 0.34~eV above H$_{BC}^{+}$. The result of the participation
of H$_{AB}^{-}$ and especially of fast diffusing H$_{AB}^{0}$ (in
addition to H$_{BC}^{+}$) in the diffusing mix, is an enhancement
of the diffusivity by a factor of ten. Of course there are large errors,
especially because at such high temperatures the harmonic approximation
is no longer applicable. However, the effect is qualitatively visible
as a bowing in the calculated diffusivity at high temperatures in
Figure~\ref{fig6}(c).

\noindent 
\begin{figure}
\includegraphics{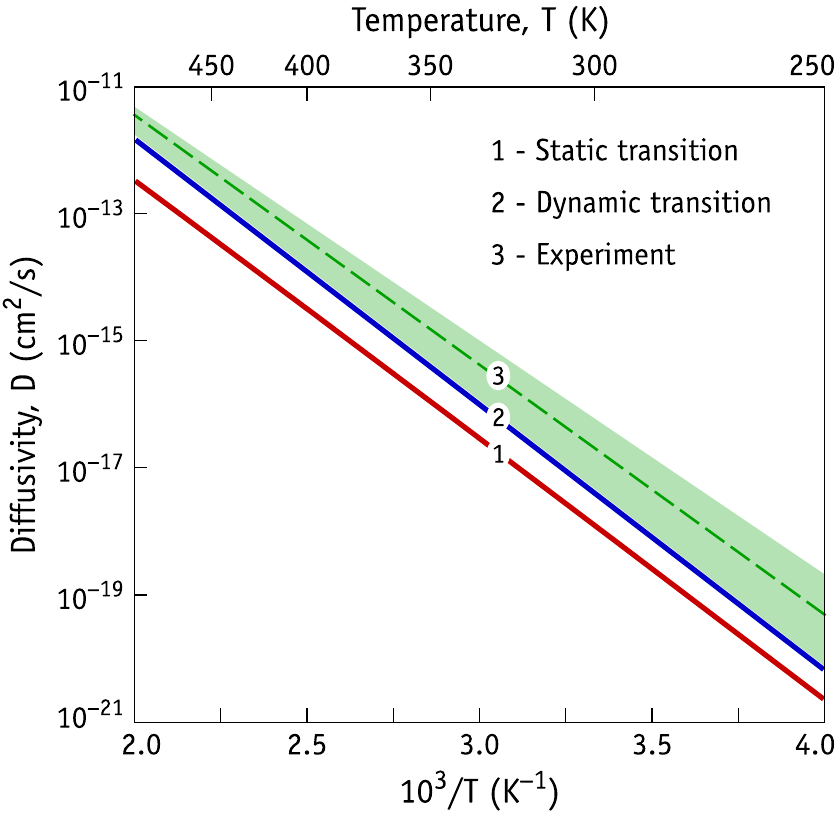}

\caption{\label{fig8} Diffusivity of molecular hydrogen in silicon. The red
and blue solid lines (1 and 2) represent the calculated diffusivity
of H$_{2T}$ assuming a static and dynamic (free rotor) molecule at
the transition state, respectively. An Arrhenius plot obtained from
measurements in the temperature range $T=300\textrm{-}420$~K \citep{Markevich1998}
is also shown (green dashed line 3). The shadow green area limits
the upper and lower bounds of the measured diffusivity of the molecule
as estimated from the error bars of the pre-exponential and energy
barrier from the fit.}
\end{figure}

We now turn our attention to the diffusivity of the hydrogen molecule
in Si. We saw in Section~\ref{subsec:migration} that the H$_{2T}$
molecule has a relatively low migration barrier ($\Delta E_{\textrm{a}}=0.82$~eV)
when compared to the barrier for conversion to H$_{2}^{*}$ ($\Delta E_{\textrm{a}}=1.62$~eV).
This implies that the molecule anneals via migration and dissociation
is most likely to occur with the action of a defect which could react
with the molecule. We also found that the variation of the potential
energy with rotation of the molecule is a few meV in both stable ($T$
site) and transition ($H$ site) states, suggesting that H$_{2T}$
migrates as a free rotor and not a static dimer aligned along the
hexagonal ring.

We calculated the attempt frequency for molecular jumping between
$T$ sites (Eq.~\ref{eq:prefactor}), assuming either a static ($Z_{\textrm{rot}}^{\ddagger}=1$)
or a dynamic transition state ($Z_{\textrm{rot}}^{\ddagger}$ given
by Eq.~\ref{eq:z_rot}). In the first case the attempt frequency
is reduced by a factor $1/Z_{\textrm{rot}}$, which accounts for the
\emph{narrowing} of the saddle point due to the absence of rotational
freedom on that state. The hindering effect is further enhanced with
raising the temperature. At $T=0$~K the reducing factor is $1/Z_{\textrm{rot}}=1$
and the attempt frequency $\nu_{0}=0.93$~THz. However, the latter
decreases by a factor of $1/Z_{\textrm{rot}}=0.26$ at $T=400$~K.
Considering that each molecule can perform up to $g_{\textrm{j}}=4$
equivalent jumps of length $l_{\textrm{j}}=2.35$~Å, we arrive at
a diffusivity represented by the red line with label 1 in Figure~\ref{fig8},
which underestimates the measured diffusivity (dashed green line with
label 3) by a factor of 10 in the temperature range $T=300\textrm{-}400$~K.

If on the other hand we consider a dynamic transition state, $\nu_{0}$
is not reduced and we arrive at a calculated diffusivity of $D=(3.4\times10^{-4}\:\textrm{cm}^{2}/\textrm{s})\:\exp(-0.82\:\textrm{eV}/k_{\textrm{B}}T)$.
This result is shown in Figure~\ref{fig8} as a solid blue line with
label 2, edging the lower bound of the experimental values (considering
their respective errors) \citep{Markevich1998}. The calculated pre-factor
of the dynamic transition mechanism overestimates the measured figure
by a factor of 1.3 only (see Table~\ref{tab1}). These results, along
with the flat potential for rotation of the molecule at the saddle
point, suggest that the H$_{2T}$ molecules not only rotate freely
on their stable sites, but they also rotate during migration.

\section{Conclusions}

We presented a theoretical study of the dynamics of atomic hydrogen
and hydrogen dimers in silicon at finite temperatures. Hybrid density
functional theory was employed to obtain the electronic (plus clamped
ionic) potential, activation barriers for migration/reconfiguration.
Semi-local density functional theory was used to calculate vibrational
frequencies within the harmonic approximation. Besides the potential
energy, we calculated free energies which account for vibrational,
rotational and configurational degrees of freedom within a diluted
regime.

For the sake of convenience, we collected all calculated properties
pertaining atomic hydrogen defects ($\textrm{H}_{BC}^{+}$, $\textrm{H}_{BC}^{0}$,
$\textrm{H}_{AB}^{0}$ and $\textrm{H}_{AB}^{-}$) and stable hydrogen
dimers (H$_{2}^{*}$ and the H$_{2T}$ molecule) in Table~\ref{tab1}.
Here one can find the configurational degeneracy per unit cell for
each defect ($g$), some bond lengths ($d$), zero-phonon energies
($\Delta E_{\textrm{ZP}}$), activation energies for migration and
reconfiguration ($\Delta E_{\textrm{a}}$), their respective attempt
frequencies ($\nu_{0}$), jumping degeneracy ($g_{\textrm{j}}$),
jump distance ($l_{\textrm{j}}$), reconfiguration and carrier emission
energies ($\Delta E$ is positive/negative for endothermic/exothermic
processes), diffusivity prefactors ($D_{0}$), and local vibrational
mode frequencies (above the Raman edge). Several quantities can be
compared directly with experimental values (rightmost column).

\noindent 
\begin{table*}
\caption{\label{tab1}Calculated properties of hydrogen monomers and dimers
in silicon. Quantities reported in the second column include: configurational
degeneracy per unit cell ($g$), bond lengths ($d$), zero-phonon
energies ($\Delta E_{\textrm{ZP}}$), activation energies ($\Delta E_{\textrm{a}}$),
attempt frequencies ($\nu_{0}$), jumping degeneracy ($g_{\textrm{j}}$),
jump distance ($l_{\textrm{j}}$), reconfiguration and carrier emission
energies ($\Delta E$ is positive/negative for endothermic/exothermic
processes), diffusivity prefactors ($D_{0}$), and local vibrational
mode (LVM) frequencies. Several measured quantities and references
can be found in the third column.}

\begin{tabular}{lcc}
\hline 
 & Calculated & Measured\tabularnewline
\hline 
$\textrm{H}_{BC}^{+}$ & $g=4$, $d_{\textrm{Si-H}}=1.592$~Å, $\Delta E_{\textrm{ZP}}=$0.14~eV & \tabularnewline
$\textrm{H}_{BC}^{+}\xrightarrow{\textrm{migration}}\textrm{H}_{BC}^{+}$ & $\Delta E_{\textrm{a}}=0.42\:\textrm{eV}$,$\nu_{0}=1.4\:\textrm{THz}$,
$g_{\textrm{j}}=6$, $l_{\textrm{j}}=1.92$~Å & $\nu_{0}=2.3\:\textrm{THz}$, $\Delta E_{\textrm{a}}=0.43\textrm{-}0.44\,\textrm{eV}$
\citep{Gorelkinskii1996,Nielsen1999}\tabularnewline
$\textrm{H}_{BC}^{+}\rightarrow\textrm{H}_{BC}^{0}+h^{+}$ & $\Delta E=0.90$~eV & \tabularnewline
LVM frequencies & 2112~cm$^{-1}$ & 1998~cm$^{-1}$ \citep{Budde2001}\tabularnewline
\hline 
$\textrm{H}_{BC}^{0}$ & $g=4$, $d_{\textrm{Si-H}}=1.614$~Å, $\Delta E_{\textrm{ZP}}=0.12$~eV & \tabularnewline
$\textrm{H}_{BC}^{0}\rightarrow\textrm{H}_{BC}^{+}+e^{+}$ & $\Delta E=0.20$~eV & $\Delta E=0.175\textrm{-}0.2$~eV \citep{Irmscher1984,Nielsen1999}\tabularnewline
$\textrm{H}_{BC}^{0}\xrightarrow{\textrm{migration}}\textrm{H}_{BC}^{0}$ & $\Delta E_{\textrm{a}}=0.36\:\textrm{eV}$,$\nu_{0}=9.4\:\textrm{THz}$,
$g_{\textrm{j}}=6$, $l_{\textrm{j}}=1.92$~Å & \tabularnewline
$\textrm{H}_{BC}^{0}\xrightarrow{\textrm{reconf.}}\textrm{H}_{AB}^{0}$ & $\Delta E=0.20$~eV, $\Delta E_{\textrm{a}}=0.36$~eV, $\nu_{0}=9.4\:\textrm{THz}$ & $\Delta E_{\textrm{a}}=0.293$~eV, $\nu_{0}=3\:\textrm{THz}$ \citep{Nielsen1999}\tabularnewline
$2\textrm{H}_{BC}^{0}\rightarrow\textrm{H}_{BC}^{+}+\textrm{H}_{AB}^{-}$ & $\Delta E=-0.21$~eV & \tabularnewline
LVM frequencies & 1998~cm$^{-1}$ & \tabularnewline
\hline 
$\textrm{H}_{AB}^{0}$ & $g=8$, $d_{\textrm{Si-H}}=1.675$~Å, $\Delta E_{\textrm{ZP}}=0.04$~eV & \tabularnewline
$\textrm{H}_{AB}^{0}\rightarrow\textrm{H}_{AB}^{-}+h^{+}$ & $\Delta E=0.49$~eV & \tabularnewline
$\textrm{H}_{AB}^{0}\xrightarrow{\textrm{migration}}\textrm{H}_{AB}^{0}$ & $\Delta E_{\textrm{a}}=0.04\:\textrm{eV}$, $\nu_{0}=5.4\:\textrm{THz}$,
$g_{\textrm{j}}=6$, $l_{\textrm{j}}=2.01$~Å & $\Delta E_{\textrm{a}}<0.1\:\textrm{eV}$ \citep{Nielsen2002}\tabularnewline
$\textrm{H}_{AB}^{0}\xrightarrow{\textrm{reconf.}}\textrm{H}_{BC}^{0}$ & $\Delta E=-0.20$~eV, $\Delta E_{\textrm{a}}=0.16\:\textrm{eV}$,
$\nu_{0}=2.1\:\textrm{THz}$ & $\Delta E_{\textrm{a}}\sim0.2$~eV \citep{Nielsen1999}\tabularnewline
LVM frequencies & 1114~cm$^{-1}$ & \tabularnewline
\hline 
$\textrm{H}_{AB}^{-}$ & $g=8$, $d_{\textrm{Si-H}}=1.771$~Å, $\Delta E_{\textrm{ZP}}=0.07$~eV & \tabularnewline
$\textrm{H}_{AB}^{-}\rightarrow\textrm{H}_{AB}^{0}+e^{-}$ & $\Delta E=0.61$~eV & $\Delta E=0.56$~eV \citep{Johnson1994}\tabularnewline
$\textrm{H}_{AB}^{-}\rightarrow\textrm{H}_{BC}^{0}+e^{-}$ & $\Delta E=0.41$~eV & $\Delta E=0.4\textrm{-}0.6$~eV \citep{Nielsen2002}\tabularnewline
$\textrm{H}_{AB}^{-}\xrightarrow{\textrm{migration}}\textrm{H}_{AB}^{-}$ & $\Delta E_{\textrm{a}}=0.47\:\textrm{eV}$, $\nu_{0}=9.6\:\textrm{THz}$,
$g_{\textrm{j}}=6$, $l_{\textrm{j}}=2.03$~Å & $\Delta E_{\textrm{a}}\lesssim0.7$~eV \citep{Johnson1992}\tabularnewline
LVM frequencies & (920, 695)~cm$^{-1}$ & \tabularnewline
\hline 
H$_{2}^{*}$ & $g=8$, $d_{\textrm{Si-H(BC)}}=1.512$~Å, $d_{\textrm{Si-H(AB)}}=1.548$~Å,
$\Delta E_{\textrm{ZP}}=0.36$~eV & \tabularnewline
$\textrm{H}_{2}^{*}\xrightarrow{\textrm{reconf.}}\textrm{H}_{2T}$ & $\Delta E=-0.19$~eV, $\Delta E_{\textrm{a}}=1.43$~eV & \tabularnewline
LVM frequencies & (2078, 1805, 776, 775)~cm$^{-1}$ & (2062, 1838)~cm$^{-1}$ \citep{Holbech1993}\tabularnewline
\hline 
H$_{2T}$ molecule & $g=2$, $d_{\textrm{H-H}}=0.778$~Å, $\Delta E_{\textrm{ZP}}=0.35$~eV & \tabularnewline
$\textrm{H}_{2T}\xrightarrow{\textrm{migration}}\textrm{H}_{2T}$ & $\Delta E_{\textrm{a}}=0.82\,\textrm{eV}$, $D_{0}=3.4\times10^{-4}\,\textrm{cm}^{2}/\textrm{s}$,
$g_{\textrm{j}}=4$, $l_{\textrm{j}}=2.35$~Å & $D_{0}=2.6\times10^{-4}\,\textrm{cm}^{2}/\textrm{s}$, $\Delta E_{\textrm{a}}=0.78\,\textrm{eV}$
\citep{Markevich1998}\tabularnewline
$\textrm{H}_{2T}\xrightarrow{\textrm{reconf.}}\textrm{H}_{2}^{*}$ & $\Delta E=0.19$~eV, $\Delta E_{\textrm{a}}=1.62$~eV, $\nu_{0}=5.4\:\textrm{THz}$ & \tabularnewline
LVM frequencies & 3687~cm$^{-1}$ & 3618~cm$^{-1}$\citep{Leitch1998,Pritchard1998}\tabularnewline
\hline 
\end{tabular}
\end{table*}

In the first part of the Results section we reproduced the well established
negative-$U$ model of atomic hydrogen in Si, involving a metastable
neutral state that disproportionates into stable bond-centered (positively
charged) or anti-bonding (negatively charged) species. The transition
between the stable states is estimated to occur when the Fermi level
crosses $E(-/+)=E_{\textrm{c}}-0.30$~eV. Metastable transition levels
at $E(0/+)=E_{\textrm{c}}-0.20$~eV and $E(-/0)=E_{\textrm{c}}-0.61$~eV
were found for the H$_{BC}$ and H$_{AB}$ species, respectively,
and they compare fairly well with the experimental findings.

The zero-point energy contribution to the calculated transition energies
is not negligible. For H$_{BC}$ this quantity differs from that of
H$_{AB}$ by about 0.1~eV. Due to cancellation effects, the impact
of zero-point motion on the calculated transition energies involving
identical configurations is minor. However, the $(-/0)$ and $(-/+)$
levels involving $\textrm{H}_{AB}^{-}\rightarrow\textrm{H}_{BC}^{0}+e^{-}$
and $\textrm{H}_{AB}^{-}\rightarrow\textrm{H}_{BC}^{+}+2e^{-}$, are
affected by errors of 0.1~eV and 0.05~eV, respectively, when zero-point
motion is not considered.

Regarding the formation mechanism of molecules in Si quenched from
high temperatures ($T\gtrsim700$~$^{\circ}$C), we found the following:

The calculated migration barrier of the molecule is about 0.8~eV.
This figure is very close to the measured value and it is consistent
with the observation of mobile molecules just above room temperature
\citep{Markevich1998}.

Based on the calculated attempt rate and barrier for dissociation
of the molecule (transformation into H$_{2}^{*}$), the annealing
temperature of H$_{2T}$ was estimated as $T\sim700$~K. This is
an upper bound as it neglects trapping by and reaction with defects.
Nevertheless, these results suggest that the lifetime of H$_{2T}$
significantly exceeds that of H$_{2}^{*}$ (which according to the
experimental results is not thermally stable above 200~$^{\circ}$C
\citep{Holbech1993}). Even up to this temperature, when formation
of H$_{2}^{*}$ can compete with that of H$_{2T}$, the molecule is
invariably more stable, and that relative stability increases with
temperature. The free energy of formation of the molecule is 0.2~eV
and 0.4~eV lower at $T=0$~K and 500~K, respectively. We could
not find any other H dimer, which could be considered stable enough
as to compete with the formation of H$_{2T}$ and H$_{2}^{*}$.

Due to configurational entropy, the stability of H monomers increases
with temperature. However, the relative stability of anti-bonding
monomers, especially the negatively charged H$_{AB}^{-}$ state, increases
faster with temperature, and therefore H$_{AB}^{-}$ becomes the second
most stable monomer after H$_{BC}^{+}$ above 500~K. We estimate
that at $T\sim1200$~$^{\circ}$C, the free energy of H$_{AB}^{-}$
is only 0.12~eV above H$_{BC}^{+}$, and its population is nearly
1/4 of the total free hydrogen.

We propose the following stages for the condensation of atomic H into
molecules during cooling: (1) above $T\sim700$~K the molecules are
not stable and free hydrogen may only be found in the atomic form,
mostly as H$_{BC}^{+}$, but also as a small fraction of H$_{AB}^{-}$;
(2) In the temperature window $T\sim700\textrm{-}500$~K the Coulomb
interaction between H$_{BC}^{+}$ and a small population of H$_{AB}^{-}$
leads to the formation stable H$_{2T}$ molecules. At this point no
other dimers are stable; (3) Below $T\sim500$~K the fraction of
H$_{AB}^{-}$ becomes negligible in comparison to other states, effectively
stopping further formation of molecules.

The above also explains why it is so difficult to form H$_{2}^{*}$
in quenched material, despite its considerable stability compared
to the molecule.

The formation of H$_{AB}^{-}$ could also partially explain the apparent
inconsistency between the high temperature and low temperature diffusivity
of atomic hydrogen in Si. While the description of the low temperature
diffusivity of the dominant $\textrm{H}{}_{BC}^{+}$ state is acceptable,
in the range $T=1300\textrm{-}1500$~K the calculated H$_{BC}^{+}$
diffusivity is lower compared to the experimentally determined values
by a factor of 20. Adding the contribution to the diffusivity from
the H$_{AB}^{-}$ population increased the effective diffusivity by
a factor of 3. However, the larger enhancement occurred when we took
into account a thermal population of fast-diffusing H$_{AB}^{0}$
species. At such high temperatures the free energy of this state is
close to that of H$_{BC}^{0}$ and their concentrations are also comparable.
However, the minute barrier for migration of H$_{AB}^{0}$ ($\Delta E_{\textrm{a}}\sim40$~meV)
increases the effective diffusivity by a factor of 10, further narrowing
the discrepancy between the calculated and experimentally determined
values.

Finally, unlike the commonly pictured mechanism for the migration
of molecular hydrogen in Si, according to which H$_{2T}$ jumps between
neighboring $T$ sites with their H-H bond perpendicular to the hexagonal
rings of the Si crystal, we found that the potential energy change
for molecular rotation at the saddle-point is only a few meV. We also
found that the observed diffusivity of the molecule is better described
if it migrates as a nearly free rotor, all along the minimum energy
path, including at the transition state.

\section*{Acknowledgments}

This work was supported by the FCT in Portugal through Projects UID-B/50025/2020,
UID-P/50025/2020, and CPCA/A0/7277/2020 (Advanced Computing Project
using the Oblivion supercomputer). The work in the UK was funded by
EPSRC via grant EP/TO25131/1. J.C. acknowledges Dr. Vladimir Voronkov
for fruitful discussions regarding the diffusivity of atomic hydrogen.

\bibliographystyle{apsrev4-1}

%

\end{document}